\documentclass[sigconf]{acmart}

\AtBeginDocument{%
  }

\copyrightyear{2024} 
\acmYear{2024} 
\setcopyright{acmlicensed}
\acmConference[ASE '24]{39th IEEE/ACM International Conference on Automated Software Engineering }{October 27-November 1, 2024}{Sacramento, CA, USA} 
\acmBooktitle{39th IEEE/ACM International Conference on Automated Software Engineering (ASE '24), October 27-November 1, 2024, Sacramento, CA, USA} 
\acmDOI{10.1145/3691620.3695506} 
\acmISBN{979-8-4007-1248-7/24/10}

\usepackage[linesnumbered,ruled,vlined]{algorithm2e}
\usepackage{amsmath}
\usepackage{makecell}
\usepackage{multirow}
\usepackage{tcolorbox}

\begin{document}
\title[A Pair Programming Framework for Code Generation]{A Pair Programming Framework for Code Generation via Multi-Plan Exploration and Feedback-Driven Refinement}

\author{Huan Zhang}
\orcid{0009-0000-0644-3775}
\affiliation{
    \department{State Key Laboratory for Novel Software Technology}
    \institution{Nanjing University} 
    \city{Nanjing} \country{China}
}
\email{zhanghuan.nju@gmail.com}

\author{Wei Cheng}
\orcid{0000-0001-6128-7293}
\affiliation{
    \department{State Key Laboratory for Novel Software Technology}
    \institution{Nanjing University} 
    \city{Nanjing} \country{China}
}
\email{wchengcs.nju@gmail.com}

\author{Yuhan Wu}
\orcid{0009-0001-9065-5523}
\affiliation{
    \department{State Key Laboratory for Novel Software Technology}
    \institution{Nanjing University} 
    \city{Nanjing} \country{China}
}
\email{yhwu.nju@gmail.com}

\author{Wei Hu}
\orcid{0000-0003-3635-6335}
\authornote{Corresponding author}
\affiliation{
    \department{State Key Laboratory for Novel Software Technology}
    \department{National Institute of Healthcare Data Science}
    \institution{Nanjing University} 
    \city{Nanjing} \country{China}
}
\email{whu@nju.edu.cn}

\renewcommand{\shortauthors}{Zhang et al.}
\newcommand{\methodname}{\textsc{PairCoder}\xspace}
\newcommand{\navigator}{\textsc{Navigator}\xspace}
\newcommand{\driver}{\textsc{Driver}\xspace}

\begin{abstract}
Large language models (LLMs) have achieved impressive performance on code generation.
Although prior studies enhanced LLMs with prompting techniques and code refinement, they still struggle with complex programming problems due to rigid solution plans.
In this paper, we draw on pair programming practices to propose \methodname, a novel LLM-based framework for code generation.
\methodname incorporates two collaborative LLM agents, namely a \navigator agent for high-level planning and a \driver agent for specific implementation.
The \navigator is responsible for proposing promising solution plans, selecting the current optimal plan, and directing the next iteration round based on execution feedback.
The \driver follows the guidance of \navigator to undertake initial code generation, code testing, and refinement.
This interleaved and iterative workflow involves multi-plan exploration and feedback-based refinement, which mimics the collaboration of pair programmers.
We evaluate \methodname with both open-source and closed-source LLMs on various code generation benchmarks.
Extensive experimental results demonstrate the superior accuracy of \methodname, achieving relative pass@1 improvements of 12.00\%--162.43\% compared to prompting LLMs directly.
\end{abstract}



\maketitle

\section{Introduction}
\label{sec:intro}

Code generation aims to automatically generate executable source code that conforms to given requirements, typically expressed in natural language. 
Recent progress in large language models (LLMs) has significantly improved software development productivity by reducing repetitive programming efforts \cite{Kazemitabaar2023Studying,Peng2023The}. 
The success of commercial models like ChatGPT \cite{chatgpt} and Claude \cite{claude}, along with powerful open-source models like Code Llama \cite{codellama} and DeepSeek Coder \cite{deepseekcoder}, has attracted substantial interest from both academia and industry. 
These advancements demonstrate the remarkable capabilities of LLMs in code generation and have great potential to influence the field of intelligent software engineering \cite{Deng2024Large,Geng2024Large,Yang2024Exploring}.

As requirements become more complex, it becomes challenging for LLMs (and even humans) to directly generate code that meets the given requirements \cite{self_collaboration}. 
One key focus of existing work is prompting techniques, which guide LLMs to produce intermediate reasoning steps for problem descriptions. 
This line of work \cite{cot,scot,self_planning} focuses on designing different types of prompts to stimulate the reasoning abilities of LLMs, enabling them to generate intermediate steps before producing the final code.
Another important aspect is that generating the correct code is rarely a one-time effort \cite{self_debugging}. 
Several studies \cite{codet,alphacode,lever} employ sampling-based approaches to filter or rank the numerous responses generated by LLMs, relying on a substantial number of samples. 
Other works \cite{self_repair,self_debugging,intervenor} attempt to refine the generated code using feedback from LLMs themselves or external sources.
They introduce a debugging process to make the generated program behave as expected.
Furthermore, a few works \cite{self_collaboration,chatdev,metagpt} explore the use of collaborative LLM agents to simulate human software development processes.

\begin{figure}
\centering
\includegraphics[width=\linewidth]{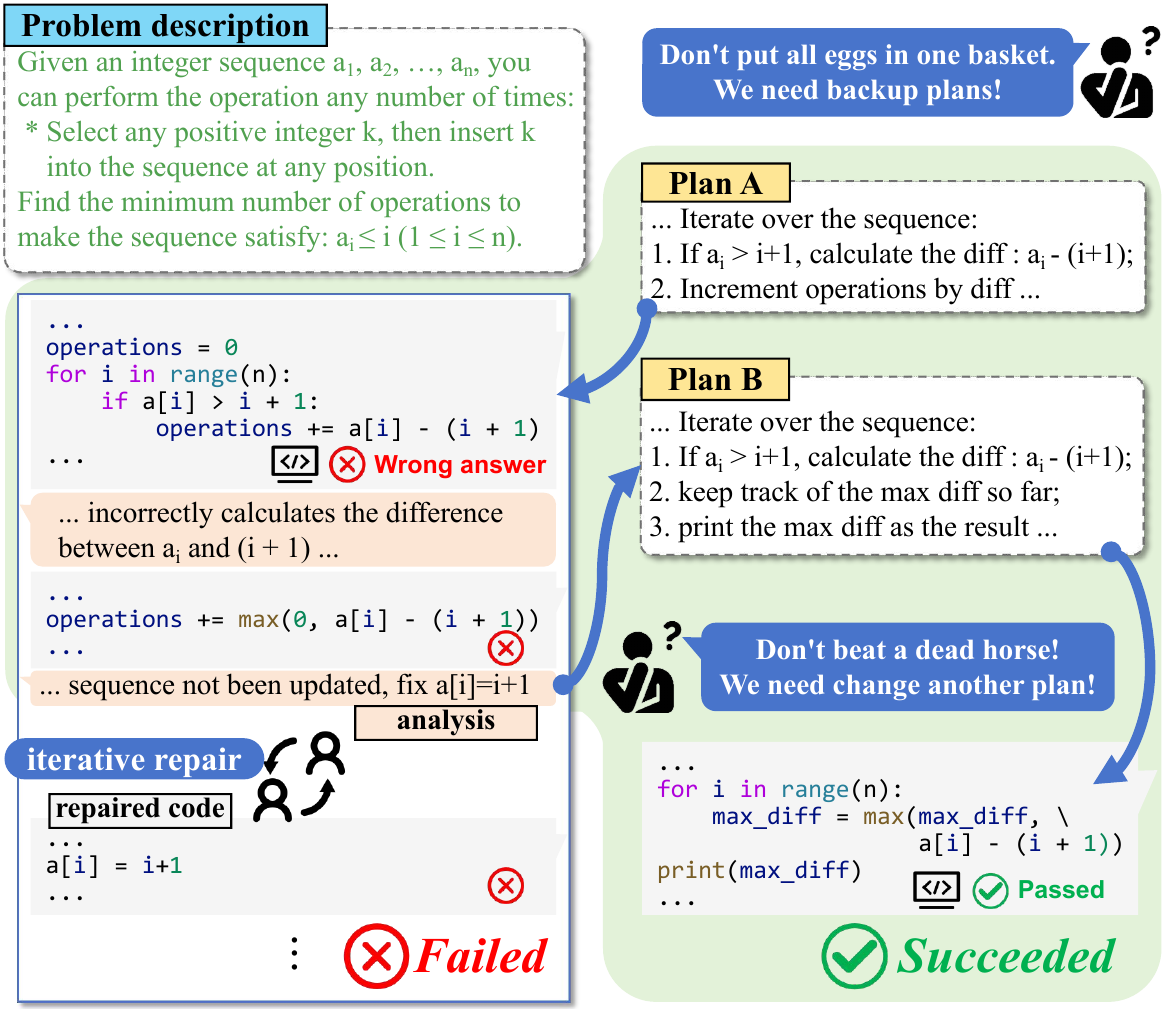}
\caption{A competitive programming problem excerpted from the CodeContest benchmark (\#test87) \cite{alphacode}. Both the plans and codes are generated by GPT-3.5-Turbo.}
\Description{A competitive programming problem excerpted from the CodeContest benchmark (\#test87) \cite{alphacode}. Both the plans and codes are generated by GPT-3.5-Turbo.}
\label{fig:example}
\end{figure}

Although existing approaches have improved code generation, they still have significant limitations.  
First, most approaches focus too narrowly on generating a single solution plan, lacking high-level planning and problem decomposition capabilities. Due to the inherent complexity of real-world problems, a single solving path often fails to produce correct code directly. 
Second, if the initially generated code follows a flawed plan, the feedback-based repair process is likely to be misguided by the initial incorrect direction, failing to address fundamental issues.
This deviates from how human developers tackle programming problems in practice. 
Human developers typically start with high-level problem analysis and propose multiple potential solution plans. 
Then, they evaluate these plans based on continuous implementation and testing feedback. 
If the current plan is deemed infeasible or reaches an impasse, they will decisively abandon it and explore alternatives. 
This iterative cycle of ``multi-plan exploration and practical feedback'' ensures that human developers can efficiently solve complex problems without being constrained by a single flawed plan.

To overcome these limitations, we propose \methodname, a novel code generation framework inspired by pair programming practices \cite{pair_programming} in software engineering. 
In pair programming, two developers play the roles of ``navigator" and ``driver": 
the former is responsible for high-level planning and direction, while the latter focuses on implementing the current task.
We adapt pair programming practices to LLM-based code generation by introducing two intelligent agents: \navigator and \driver. 
The \navigator's role includes starting with high-level problem analysis (referred to as reflection), generating multiple potential solution plans, selecting the best current plan, and directing the subsequent process based on the \driver's execution feedback. 
The \driver generates or repairs the code according to the \navigator's guidance and provides execution feedback for subsequent adjustment.

This \navigator-\driver framework mimics the iterative and adaptive strategies employed by human developers. 
By starting with reflection and generating multiple plans, it avoids the constraints of a single flawed plan. 
Continuous feedback interaction between the \navigator and the \driver allows for dynamic plan adjustment, enabling the abandonment of infeasible plans and exploring new directions when necessary. 
This iterative cycle significantly improves the robustness and quality of code generation for real-world tasks.
We conduct extensive experiments to evaluate our \methodname on five diverse code generation benchmarks using both open-source (DeepSeek-Coder \cite{deepseekcoder}) and closed-source (GPT-3.5-Turbo \cite{gpt4_report}) LLMs. 
The results show that on the pass@1 metric, \methodname achieves superior accuracy over competitive baselines across all benchmarks for both LLMs.

In summary, the key contributions in this paper are outlined as follows:
\begin{itemize}
\item We are the first to adapt pair programming practices into LLM-based code generation and propose a new framework called \methodname that comprises two collaborative agents: a \navigator for high-level planning and a \driver for specific implementation. (Sect.~\ref{sec:overview})
\item \methodname integrates two key mechanisms: (i) multi-plan exploration achieved by the \navigator through adjusting diverse solution plans, and (ii) feedback-driven refinement based on execution feedback from the \driver and historical memory. (Sects.~\ref{sec:navigator} and \ref{sec:driver})
\item \methodname significantly outperforms all competitive baselines on five benchmarks. 
It gains relative improvements of 16.97\%--162.43\% on GPT-3.5-Turbo and 12.00\%--128.76\% on DeepSeek-Coder-Instruct 33B compared to prompting LLMs directly. (Sect.~\ref{sec:exp})
The source code is publicly available on GitHub.\footnote{\url{https://github.com/nju-websoft/PairCoder}}
\end{itemize}

\section{Motivating Example}
\label{sec:example}

Fig.~\ref{fig:example} shows a programming problem from the CodeContest benchmark \cite{alphacode}: given an integer sequence $a_1, a_2, \dots, a_n$, find the minimum number of insertions to ensure $\forall 1 \leq i \leq n,  a_i \leq i$.

The prompting approaches \cite{cot,scot,self_planning} generate intermediate reasoning steps based on the problem description to guide code generation, as excerpted in Plan A.
Considering an input sequence ``1, 3, 4" in public test cases, it is evident that a single operation is sufficient to satisfy the requirement, i.e., inserting `2' between `1' and `3'.
Guided by Plan A, the first code snippet on the left side of Fig.~\ref{fig:example} incorrectly assumes that two insertions are required.
This feedback from code testing can help LLMs repair the generated code \cite{self_debugging,self_repair,intervenor}.
However, iterative repairs do not yield the right answers as they stubbornly follow the flawed Plan A, which incorrectly accumulates \texttt{operations}.
For example, LLMs would mistake the bug as an incorrect calculation of the difference between $a_i$ and $(i + 1)$, or that the sequence has not been updated.
We believe this dilemma arises from an inherent pitfall of the single-path approach: once an incorrect blueprint is initially established, LLMs struggle to identify the root cause of the error and thus mislead subsequent repairs.

\begin{figure*}[htbp]
\centering
\includegraphics[width=0.9\linewidth]{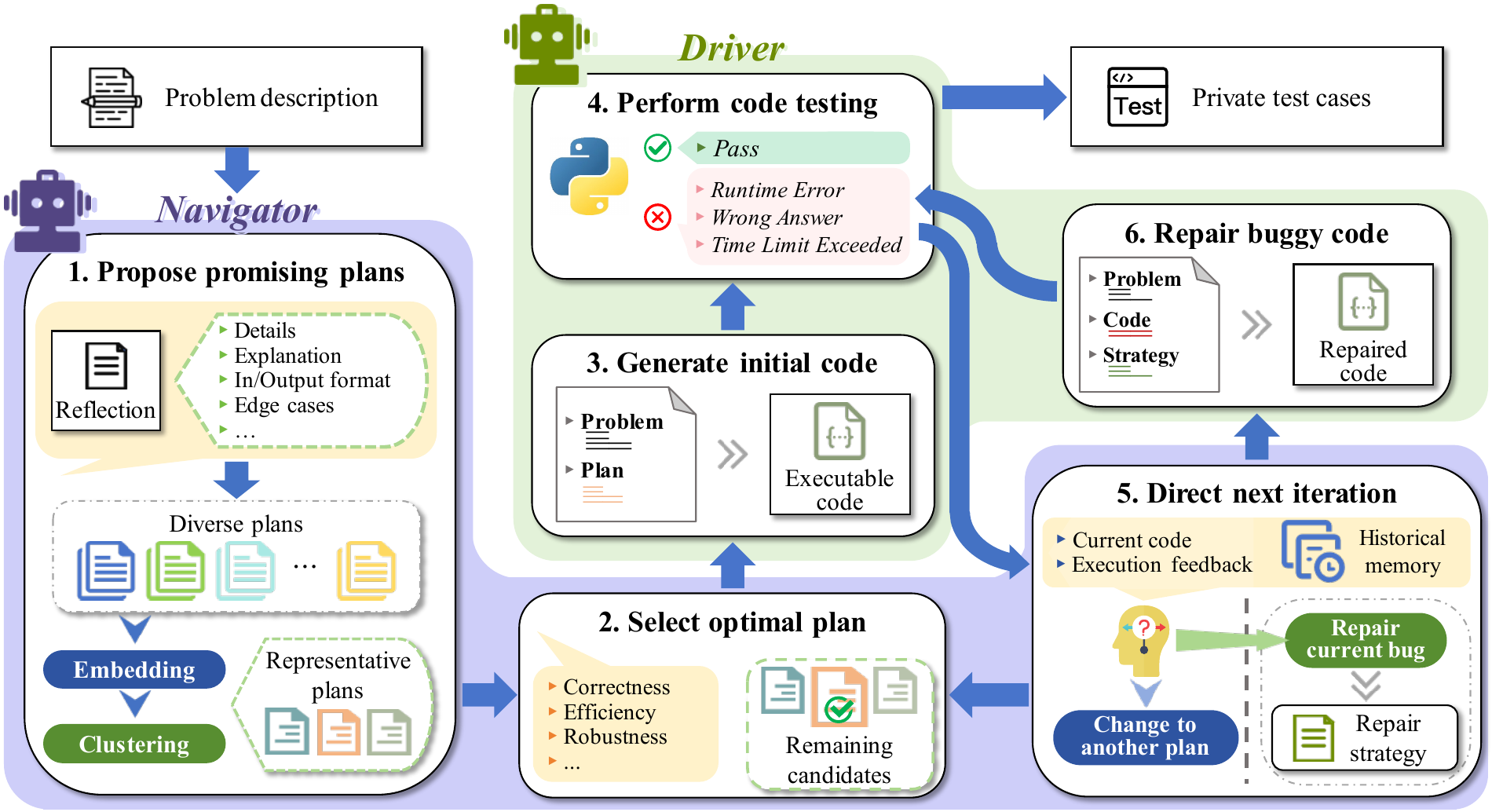}
\caption{Overview of our \methodname, in which a \navigator agent and a \driver agent collaborate on code generation.}
\Description{Overview of our \methodname, in which a \navigator agent and a \driver agent collaborate on code generation.}
\label{fig:approach}
\end{figure*}

In contrast, human programmers do not put all eggs in one basket.
If the current plan is deemed ineffective, they will explore alternative plans.
Plan B correctly solves the problem by tracking the maximum difference between $a_i$ and $(i + 1)$.
With the proper solution plan, LLMs generate the correct code instead of getting stuck in a refinement loop under Plan A.

In practice, solving complex problems often requires exploring multiple solution plans, and timely adjustment can mitigate wasted effort on flawed plans.
The role of multi-plan exploration and flexible adjustment is analogous to that of the ``navigator-driver" in pair programming, serving as the key inspiration for our work.

\section{Framework}
\subsection{Overview}
\label{sec:overview}

We first formulate the realistic problem of code generation as generating a program $C$ from a natural language description $\mathcal{Q}$ and a set of public (visible) test cases $\mathcal{T}_v = \big\{(I_i, O_i)\big\}_{i=1}^{m_v}$, where $O_i$ denotes the desired output for the input $I_i$.
Based on the execution feedback from $\mathcal{T}_v$, LLMs can iteratively refine the generated program $C$, having a maximum number of iterations $r$ to control cost and efficiency.
Finally, $C$ is considered correct if its behavior is consistent with the test oracle \cite{Howden1978Theoretical}, which is usually represented by a set of private (hidden) test cases $\mathcal{T}_h = \big\{(I_i, O_i)\big\}_{i=1}^{m_h}$, i.e., satisfying that $\forall\, (I_i, O_i) \in \mathcal{T}_h, C(I_i) = O_i$.
Note that the accessible $\mathcal{T}_v$ is not a complete test, while $\mathcal{T}_h$ is invisible during the code generation and refinement stages.

Fig.~\ref{fig:approach} illustrates the workflow of \methodname.
Both the \navigator and \driver agents are powered by an LLM with general-purpose capabilities, such as GPT-3.5-Turbo \cite{chatgpt}.
The \navigator guides the \driver by generating solution plans and repair strategies.
Therefore, the \driver focuses all its attention on specific code tasks, including code generation, code testing, and refinement.
The workflow of the two agents is interleaved and iterates until the generated program $C$ passes all public test cases $\mathcal{T}_v$ or the maximum number of iterations $r$ is reached.
To clearly describe the details of \methodname, we present Algorithm \ref{algo:pairCoder} throughout this section.

\begin{algorithm}[!t]
\caption{\methodname}
\label{algo:pairCoder}
\KwIn{
    problem description $\mathcal{Q}$, public test cases $\mathcal{T}_{v}$ \\
    [\textit{Agents}]: \textsc{Navigator} model $\mathcal{M}_N$, \textsc{Driver} model $\mathcal{M}_D$ \\
    [\textit{Memories}]: code history $\mathcal{H}_c$, feedback history $\mathcal{H}_f$ \\
    [\textit{Hyperparameters}]: the maximum number of iterations $r$, \\
    the number of sampled plans $n$, the number of clusters $k$
}

\KwOut{generated code $C$}

$reflection \gets \mathcal{M}_N(\cdot\,|\,\texttt{ReflectPrompt}, \mathcal{Q})$\;

$\{{P}_i\}_{i=1}^n \gets \mathcal{M}_N(\cdot\,|\,\texttt{PlanPrompt}, \mathcal{Q}, reflection)$\;

$S \gets \textsc{ClusterPlans}(\{{P}_i\}_{i=1}^n, k)$\tcp*[l]{\small $S \subseteq \{{P}_i\}_{i=1}^n \wedge \lVert S \rVert = k$}

$C \gets NULL$; $F \gets NULL$\;

\For{$j := 1$ \KwTo $r$}{
    \If{$(j = 1) \vee (C \in \mathcal{H}_c) \vee (F \in \mathcal{H}_f)$}{
        \tcp{\small select a new solution plan}
        $plan \gets \mathcal{M}_N(\cdot\,|\,\texttt{SelectPrompt}, \mathcal{Q}, reflection, S)$\;
        
        $S \gets S \,\backslash\, \{plan\}$\;
        $\mathcal{H}_c \gets \emptyset$;  $\mathcal{H}_f \gets \emptyset$\;
        
        $C \gets \mathcal{M}_D(\cdot\,|\,\texttt{CodePrompt}, \mathcal{Q}, plan)$\;
    }
    \Else(\tcp*[h]{\small stick to the current plan}){
        $\mathcal{H}_c \gets \mathcal{H}_c \cup \{C\}$;  $\mathcal{H}_f \gets \mathcal{H}_f \cup \{F\}$\;  
        $strategy \gets \mathcal{M}_N(\cdot\,|\,\texttt{AnalyzePrompt}, \mathcal{Q}, C, F)$\;
        
        $C \gets \mathcal{M}_D(\cdot\,|\,\texttt{RepairPrompt}, \mathcal{Q}, C, strategy)$\;
    }
    
    $F \gets \textsc{ExecuteTests}(C, \mathcal{T}_{v})$\;
    
    \lIf{$F = \textit{Pass}$}{
        \textbf{break}
    }
}
\KwRet $C$\;

\end{algorithm}


\subsection{\navigator Agent}
\label{sec:navigator}

The \navigator agent serves as the main controller in deeply understanding the problem and providing strategic direction.
Its role is to propose multiple promising plans (Step 1 in Fig.~\ref{fig:approach}), select the currently best solution plan (Step 2), and direct the next iteration based on execution feedback and historical memory (Step 5).

\begin{figure}
\centering
\includegraphics[width=\linewidth]{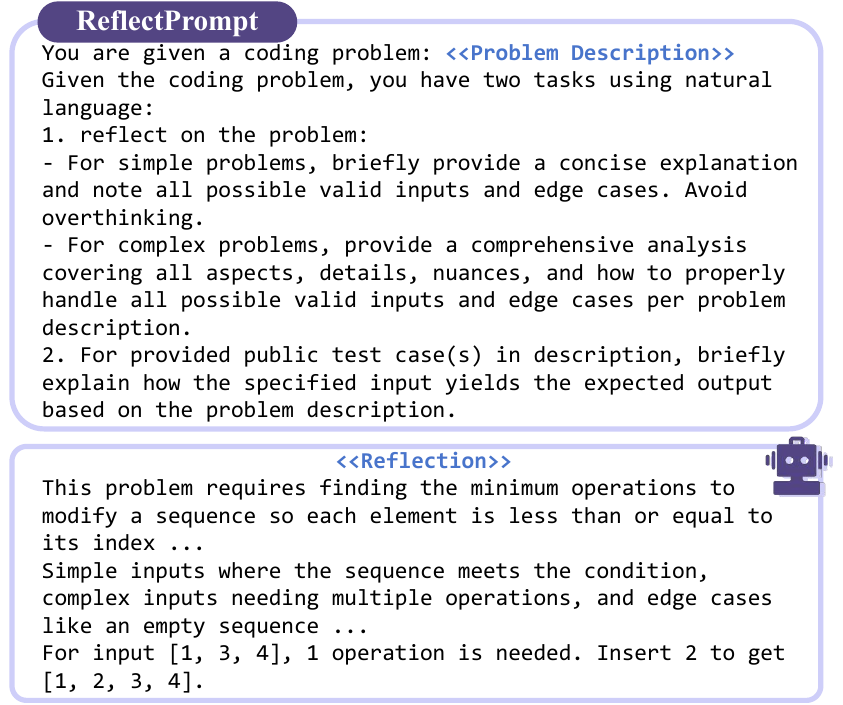}
\caption{The prompt template used by the \navigator to reflect on the given problem description and an output example of the reflection.}
\Description{The prompt template used by the \navigator to reflect on the given problem description and an output example of the reflection.}
\label{fig:reflectprompt}
\end{figure}

\paragraph{Propose promising plans}
The \navigator first reflects on the given natural language description $\mathcal{Q}$ (Line 1 in Algorithm~\ref{algo:pairCoder}).
The prompt for LLMs is shown in Fig.~\ref{fig:reflectprompt}.
It stimulates LLMs to explicitly analyze the details of the problem, consider possible valid inputs and edge cases, and explain public test cases.
This reflection process enables the \navigator to gain a comprehensive understanding of the core logic, constraints, and requirements for effective problem solving.

\begin{figure}
\centering
\includegraphics[width=\linewidth]{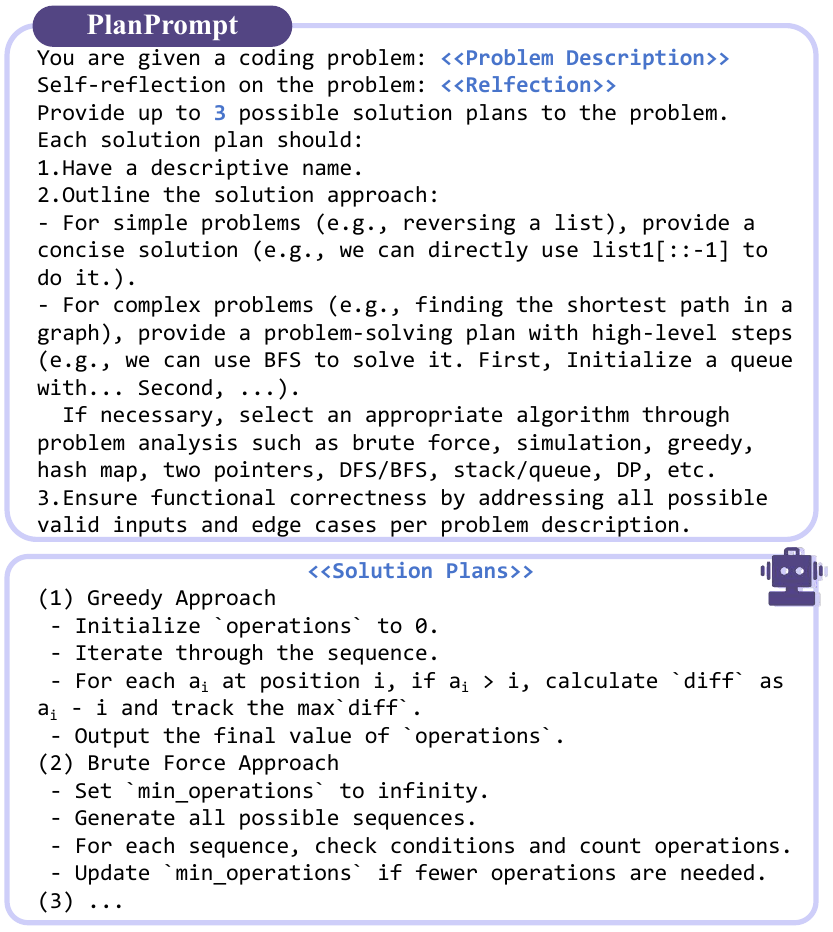}
\caption{The prompt template for sampling multiple plans and an output example of solution plans.}
\Description{The prompt template for sampling multiple plans and an output example of solution plans.}
\label{fig:planprompt}
\end{figure}

With the comprehensive reflection on the problem, the \navigator further comes up with specific solution plans (Line 2).
As shown in Fig.~\ref{fig:planprompt}, we include brief examples in prompts to guide the proposal and emphasize the functional correctness of the proposed plans.
Each plan outlines a high-level solution and key implementation steps in concise natural language.
To obtain diverse solution plans, we set a non-zero temperature for multiple nucleus sampling \cite{Holtzman2020The} and ask LLMs to generate multiple plans in each batch \cite{batch_prompting}, which improves sampling efficiency while reducing duplicate plans.
After brainstorming through multiple sampling, we select $k$ representative plans as candidates.
Specifically, we first divide all $n$ samples into $k$ clusters using a text embedding model and the classical k-means++ algorithm \cite{kmeans}, and then select the plan closest to the cluster centroid from each cluster (Line 3).
Intuitively, the \navigator groups similar plans together and selects representative ones, ensuring a diverse set of high-level solution plans. These can involve techniques like brute force search, greedy algorithm, or other problem-solving strategies.

\begin{figure}
\centering
\includegraphics[width=\linewidth]{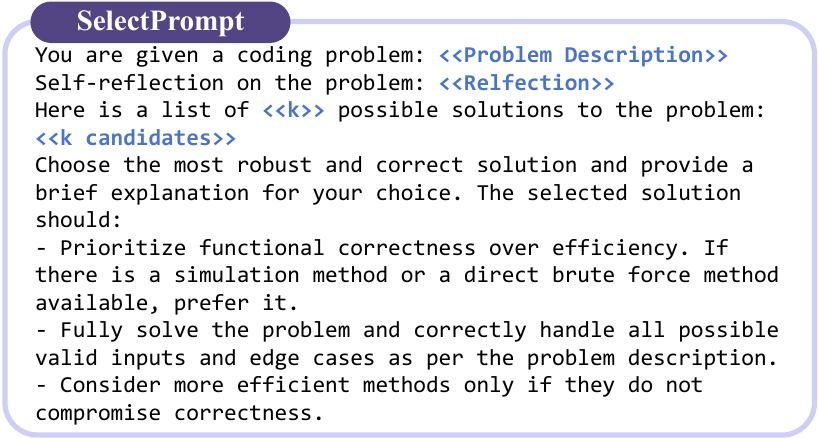}
\caption{The prompt template for selecting the optimal solution plan from the candidates.}
\Description{The prompt template for selecting the optimal solution plan from the candidates.}
\label{fig:selectprompt}
\end{figure}

\paragraph{Select optimal plan}
In each iteration, there is a determined solution plan to guide code generation or refinement.
Whenever a plan needs to be initially selected or discarded, the \navigator selects the optimal plan from the remaining candidates (Line 7).
Following the previous studies \cite{coderanker, codereviewer}, we also make choices through LLMs, where the prompt template is shown in Fig.~\ref{fig:selectprompt}.
Considering the problem description and reflection together, the \navigator leverages the reasoning capabilities of LLMs to consolidate multiple key factors in code quality, including correctness, efficiency, and robustness.
Note that functional correctness takes precedence over efficiency.
We prefer to try an inefficient but intuitive solution plan first, such as a direct brute force method, which is in line with the common practice of human programmers to prioritize problem-solving before optimization.
Once a plan is selected, it is removed from the candidates, preventing it from being selected again in subsequent iterations (Line 8).

\paragraph{Direct next iteration}
Once the generated code $C$ in the last iteration does not pass all the public test cases $\mathcal{T}_v$, it is the \navigator's turn to direct the next iteration.
Instead of stubbornly persisting in a single solving path to repair the incorrect code \cite{self_debugging, intervenor, ldb, reflexion}, the \navigator can timely adjust the solution plan to seek a turnaround.
We observe that code refinement tends to get stuck in a dead-end loop if the generated code or execution feedback has already occurred in the past.
Therefore, we introduce a long-term memory module to systematically store and maintain the coding and execution history under the current solution plan (Line 12).
It consists of the generated programs $\mathcal{H}_c = \{C^i\}_{i=1}^{r}$ and the execution feedback $\mathcal{H}_f = \{F^i\}_{i=1}^{r}$, where $F^i$ is the execution feedback of the generated code $C^i$ in the $i$-th round of iteration, comprising specific error type and execution details.
We apply a simple but effective \emph{heuristic strategy} to determine whether to change the solution plan.
Given the buggy code and its execution feedback, the current solution plan will be considered unpromising if any of them has already occurred in the historical memory $\mathcal{H}_c$ and $\mathcal{H}_f$ (Line 6), leading to a re-selection of the optimal plan in Step 2.
Whenever a new solution plan is selected, the historical memory module is cleared to start fresh (Line 9).

\begin{figure}
\centering
\includegraphics[width=\linewidth]{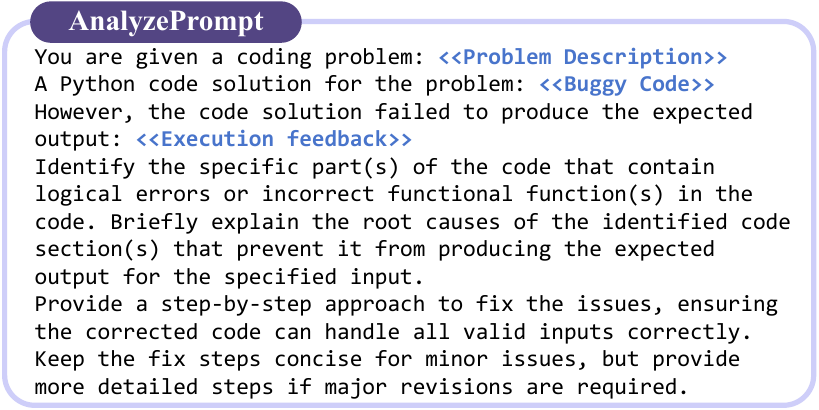}
\caption{The prompt template used to propose a repair strategy. This is customized for \textit{Wrong Answer} and is slightly different from the other types of execution feedback.}
\Description{The prompt template used to propose a repair strategy. This is customized for \textit{Wrong Answer} and is slightly different from the other types of execution feedback.}
\label{fig:analyzeprompt}
\end{figure}

Another potential iteration direction is to repair the buggy code, which pursues gradual progress without abandoning a promising solution plan.
Based on the execution feedback, the \navigator leverages the reasoning ability of LLMs to propose a directive repair strategy (Line 13).
As shown in Fig.~\ref{fig:analyzeprompt}, the prompt for LLMs comprises the problem description, the buggy code, and specific execution feedback.
For different types of execution feedback, we slightly adjust the prompts to remind LLMs to focus on specific aspects.
Specifically, \textit{Runtime Error} can suggest repairs that involve syntax error or exception handling, e.g., array index out of bounds.
\textit{Wrong Answer} indicates logic errors in the code, which may inspire repair strategies like adjusting condition handling and data flow.
For the inefficient solution plans flagged by \textit{Time Limit Exceeded}, the \navigator may recommend optimizations to improve computational performance.
The customized repair strategy will guide the code refinement in Step 6 of the next iteration.

\subsection{\driver Agent}
\label{sec:driver}

In contrast to the high-level planning of the \navigator, the \driver agent focuses all its attention on specific code tasks, including generating initial code guided by a new plan (Step 3), testing code on public test cases (Step 4), and repairing the buggy code (Step 6).

\begin{figure}
\centering
\includegraphics[width=\linewidth]{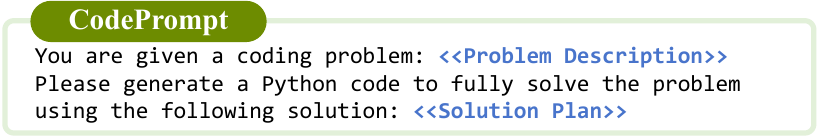}
\caption{The prompt template used by the \driver to generate initial code guided by a new solution plan.}
\Description{The prompt template used by the \driver to generate initial code guided by a new solution plan.}
\label{fig:codeprompt}
\end{figure}

\paragraph{Generate initial code}
Once a new solution plan is selected, the \driver first generates an initial code implementation guided by the plan (Line 10). 
To align with the high-level planning of the \navigator, the \driver concatenates the problem description and current solution plan together, making the prompt shown in Fig.~\ref{fig:codeprompt}.
Existing LLMs \cite{chatgpt, deepseekcoder} already have impressive abilities to comprehend the context in prompts and convert solution plans expressed in natural language into corresponding executable programs \cite{cot,brainstorm,self_planning}.
We expect, but do not mandate, that the generated code is an exact match to the solution plan. 
Incorrect implementation is tolerated at this step, which would be recognized and refined in subsequent iterations.

\paragraph{Perform code testing}
Generating correct code is rarely a one-time effort, and recent studies indicate that LLMs struggle to self-correct their responses without external feedback \cite{Valmeekam2023Can, Huang2024Large}.
Therefore, we follow previous works \cite{self_debugging, self_repair} to introduce an executor that evaluates the generated program $C$ on the public test cases $\mathcal{T}_v$ (Line 15).
We categorize the execution feedback into four types:

\begin{itemize}
    \item \textit{Pass.} It indicates that $C$ successfully passes all public test cases, i.e., satisfying that $\forall\, (I_i, O_i) \in \mathcal{T}_v, C(I_i) = O_i$.
    \item \textit{Runtime Error.} It indicates that the execution is terminated prematurely due to unhandled exceptions or errors.
    \item \textit{Wrong Answer.} It indicates that $C$ gives unexpected outputs in some cases, i.e., satisfying that $\exists\, (I_i, O_i) \in \mathcal{T}_v, C(I_i) \neq O_i$. This type takes precedence over \textit{Time Limit Exceeded}.
    \item \textit{Time Limit Exceeded.} It indicates that $C$ fails to produce outputs within the specified time limit.
\end{itemize}

If the execution feedback is \textit{Pass}, we will terminate the iterative process and consider $C$ as the final output (Line 16);
Otherwise, the \driver will deliver the current program and execution feedback to the \navigator, which are used to direct the next iteration in Step 5.
Code testing marks the end of an iteration round, so the iterative process will also be terminated after performing the $r$-th test, outputting the last generated code (Line 17).

\begin{figure}
\centering
\includegraphics[width=\linewidth]{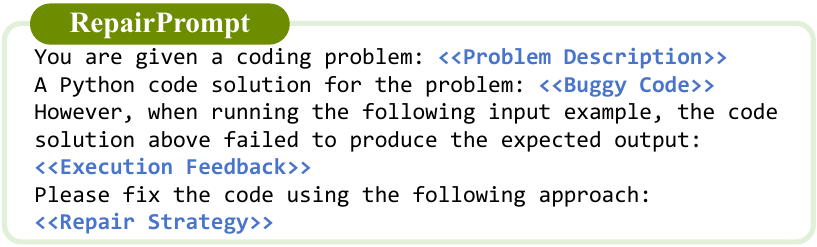}
\caption{The prompt template for repairing buggy code.}
\Description{The prompt template for repairing buggy code.}
\label{fig:repairprompt}
\end{figure}

\paragraph{Repair buggy code}
If the \navigator assumes that the current solution plan remains promising, the \driver will attempt to repair the buggy code based on the given repair strategy (Line 14).
As shown in Fig.~\ref{fig:repairprompt}, the prompt for LLMs comprises the problem description, the buggy code, the execution feedback, and the repair strategy.
The \driver aims to address the issues identified in the repair strategy, such as logic errors and performance bottlenecks.
Similar to the code generation in Step 3, we do not claim that the repair is a complete success, and the generated code can be further refined in subsequent iterations.

\paragraph{Complexity analysis} 
The time complexity of Algorithm~\ref{algo:pairCoder} is determined by the number of iterations $r$ and the cost of operations within each iteration. Let $c$ denote the constant factor representing the cost of operations, such as model inference and code testing within each iteration. Then, the overall time complexity of \methodname is $O(r \times c)$. Similarly, the space complexity of \methodname is $O(r)$, since the \navigator needs to store the historical memory $\mathcal{H}_c$ and $\mathcal{H}_f$, which grow linearly with the iteration count.
While the multi-plan exploration and iterative refinement introduce additional computational overhead, the superior accuracy of \methodname in code generation justifies the trade-off in Sect.~\ref{sec:exp}.

\section{Experiments and Results}
\label{sec:exp}

We evaluate \methodname by defining the following research questions (RQs) and outlining how we propose to answer them:
\begin{itemize}
    \item \textbf{RQ1. How does the accuracy of \methodname in code generation compare to other approaches?} We aim to evaluate the effectiveness of our \methodname framework in code generation compared to existing approaches. We conduct comprehensive experiments across diverse benchmarks and foundation models.
    \item \textbf{RQ2. How do critical hyperparameters impact the accuracy of \methodname?} 
    We thoroughly investigate the effect of iteration count on \methodname compared to other iterative refinement-based approaches, as well as the impact of cluster number on \methodname.
    \item \textbf{RQ3. What are the individual contributions of the major components in \methodname?} 
    We aim to analyze the effectiveness of two major components in \methodname: multi-plan exploration and feedback-driven refinement facilitated by \navigator-\driver collaboration. By disabling each component in ablation studies, we isolate their effects and validate their contributions to the overall accuracy.
    \item \textbf{RQ4. What are the findings of cost and error analyses for \methodname?} 
    The cost analysis focuses on quantifying the API calls and token consumption, providing insights into the computational resources required to deploy \methodname in real-world scenarios.
    We also analyze the causes of errors in failed test cases, indicating potential future improvements in code generation.
\end{itemize}

\subsection{Experiment Settings}
\label{sec:exp_settings}

\paragraph{Benchmarks}
Following the prior works \cite{self_debugging,intervenor}, we conduct comprehensive experiments on five widely used benchmarks of code generation: HumanEval \cite{humaneval}, HumanEval+ \cite{evalplus}, MBPP \cite{mbpp}, MBPP+ \cite{evalplus}, and CodeContest \cite{alphacode}.
The statistics of these benchmarks are shown in Table \ref{tab:benchmark_statistics}.
HumanEval, MBPP, and their extended versions (Plus) aim at simple function-level code generation, while CodeContest consists of competition-level programming problems.
Both the validation and test sets of CodeContest are considered.

Furthermore, we put effort into providing public test cases $\mathcal{T}_v$ for execution feedback.
For the benchmarks \cite{humaneval, evalplus, alphacode} where public test cases are provided in problem descriptions, we extract $\mathcal{T}_v$ using hand-written rules.
For the benchmarks lacking public test cases in the descriptions, we follow \cite{self_debugging, ldb, lever} by treating the first private case as $\mathcal{T}_v$ for MBPP, while for MBPP+, we use the original three private cases before extension as $\mathcal{T}_v$.

\paragraph{Metrics} 
In line with previous works \cite{self_debugging, self_collaboration, self_planning, clarifygpt}, we use the greedy pass@1 \cite{humaneval,Yu2024CoderEval} to assess the functional correctness of the generated program.
A program is regarded correct only if it passes all private test cases $T_h$.
Compared to pass@K with multiple nucleus sampling, the greedy pass@1 represents a more realistic scenario, where developers are not required to review the correct one from multiple solutions.

\begin{table}[!t]
    \centering
    \caption{Benchmark statistics, including the number of programming problems $\mathcal{Q}$, the average number of public test cases $m_v$, and the average number of private test cases $m_h$.}
    \label{tab:benchmark_statistics}
    {\small
    \begin{tabular}{l|rrrrrr}
        \toprule
        \multirow{2}{*}{Features} & \multicolumn{2}{c}{HumanEval} & \multicolumn{2}{c}{MBPP} & \multicolumn{2}{c}{CodeContest} \\
        \cmidrule(lr){2-3} \cmidrule(lr){4-5} \cmidrule(lr){6-7}
        & Orig & Plus & Orig & Plus & Valid & Test \\
        \midrule
        \# $\mathcal{Q}$ & 164 & 164 & 500 & 399 & 117 & 165 \\
        Avg. $m_v$ & 2.8 & 2.8 & 1.0 & 3.0 & 1.5 & 1.7 \\
        Avg. $m_h$ & 9.6 & 764.1 & 3.1 & 108.5 & 202.9 & 202.1 \\
        \bottomrule
    \end{tabular}}
\end{table}

\paragraph{Comparative methods} 
We compare \methodname with two main categories of existing approaches for code generation.
We briefly describe them as follows.

\textbf{Prompting techniques.} This category focuses on prompts to steer LLMs towards generating more accurate code solutions for requirements. 
Notable approaches include:

\begin{itemize}
\item \textbf{Direct prompting} \cite{humaneval} takes the original requirements directly as inputs to prompt LLMs for code generation.
\item \textbf{Chain-of-Thought (CoT) prompting} \cite{cot} elicits LLMs to generate a chain of intermediate natural language reasoning steps before producing the final code. We use the classical CoT instruction ``Let's think step by step." to guide LLMs in zero-shot \cite{cot_zero}.
\item \textbf{SCoT prompting} \cite{scot} asks LLMs using three basic program structures (i.e., sequence, branch, and loop structures) to build intermediate reasoning steps. Then, LLMs generate the final code based on the structured CoT. 
\item \textbf{Self-planning} \cite{self_planning} consists of planning and implementation phases. In the planning phase, LLMs plan out the solution steps from the intent with few-shot demonstrations. In the implementation phase, LLMs generate code step by step, guided by the planned solution steps. 
\end{itemize}

\begin{table*}
\caption{Comparison of pass@1 with two LLMs on code generation benchmarks. ``\textsuperscript{\textdagger}'' denotes the value is directly cited from the respective original work, and ``-'' denotes the empty result due to reproducibility issues.}
\label{tab:exp-main}
\centering
{\small
\begin{tabular}{l|rrrrrr|rrrrrr}
\toprule
\multirow{3}{*}{Approaches} & \multicolumn{6}{c|}{GPT-3.5-Turbo} & \multicolumn{6}{c}{DeepSeek-Coder} \\
\cmidrule(lr){2-7} \cmidrule(lr){8-13}
& \multicolumn{2}{c}{HumanEval} & \multicolumn{2}{c}{MBPP} & \multicolumn{2}{c|}{CodeContest} & \multicolumn{2}{c}{HumanEval} & \multicolumn{2}{c}{MBPP} & \multicolumn{2}{c}{CodeContest} \\
\cmidrule(lr){2-3} \cmidrule(lr){4-5} \cmidrule(lr){6-7} \cmidrule(lr){8-9} \cmidrule(lr){10-11} \cmidrule(lr){12-13}
& Orig & Plus & Orig & Plus & Valid & Test & Orig & Plus & Orig & Plus & Valid & Test \\
\midrule
Direct prompting & 67.68 & 60.98 & 66.80 & 66.42 & 6.84 & 6.06 & 76.22 & 67.78 & 66.40 & 64.41 & 5.98 & 6.67 \\
CoT prompting & 68.90 & 62.80 & 69.00 & 67.17 & 5.13 & 5.45 & 77.27 & 68.90 & 67.60 & 67.67 & 5.45 & 6.67 \\ 
SCoT prompting & 68.29 & 61.59 & 62.60 & 61.40 & 5.98 & 4.24 & 73.17 & 65.85 & 61.80 & 60.15 & 4.27 & 6.06 \\
Self-planning & 72.56 & 64.63 & 69.60 & 67.67 & 7.69 & 6.06 & 74.39 & 68.90 & 65.80 & 68.17 & 6.84 & 9.09 \\
\midrule
Self-collaboration\textsuperscript{\textdagger} & 74.40 & - \ \ \ \ & 68.20 & - \ \ \ \ & - \ \ \ \ & - \ \ \ \ & - \ \ \ \ & - \ \ \ \ & - \ \ \ \ & - \ \ \ \ & - \ \ \ \ & - \ \ \ \ \\
Self-repair & 73.17 & 65.24 & 70.60 & 69.42 & 7.69 & 6.67 & 77.44 & 70.12 & 70.00 & 70.43 & 5.98 & 7.27 \\
Self-debugging & 78.05 & 72.56 & 72.80 & 70.43 & 9.40 & 11.52 & 79.27 & 73.78 & 72.20 & 71.12 & 8.55 & 13.33 \\
INTERVENOR & 77.44 & 69.51 & 73.40 & 71.93 & 8.55 & 9.09 & 79.88 & 72.56 & 72.60 & 72.43 & 7.69 & 10.30 \\
Reflexion & 69.57 & - \ \ \ \ & 67.76 & - \ \ \ \ & - \ \ \ \ & - \ \ \ \ & 81.99 & - \ \ \ \ & 74.31 & - \ \ \ \ & - \ \ \ \ & - \ \ \ \ \\
LDB\textsuperscript{\textdagger} & 82.90 & - \ \ \ \ & 76.00 & - \ \ \ \ & - \ \ \ \ & - \ \ \ \ & - \ \ \ \ & - \ \ \ \ & - \ \ \ \ & - \ \ \ \ & - \ \ \ \ & - \ \ \ \ \\
\midrule
\methodname (ours) & \textbf{87.80} & \textbf{77.44} & \textbf{80.60} & \textbf{77.69} & \textbf{17.95} & \textbf{15.15} & \textbf{85.37} & \textbf{76.22} & \textbf{78.80} & \textbf{75.69} & \textbf{13.68} & \textbf{14.55} \\
\midrule
Relative improvement $\uparrow$ & 29.73\% & 26.99\% & 20.66\% & 16.97\% & 162.43\% & 150.00\% & 12.00\% & 12.45\% & 18.67\% & 17.51\% & 128.76\% & 118.14\% \\
\bottomrule
\end{tabular}}
\end{table*}

\begin{table}
\centering
\caption{Additional comparison of pass@1 with GPT-4 on the original HumanEval and the sanitized MBPP \cite{mbpp}.}
\label{tab:exp-gpt4}
{\small
\begin{tabular}{l|cc}
\toprule
Approaches & HumanEval & MBPP \\
\midrule
Direct prompting & 84.76 & 82.71 \\
\midrule
MetaGPT\textsuperscript{\textdagger} & 85.9 & 87.7 \\
Reflexion\textsuperscript{\textdagger} & 91.0 & 77.1 \\
Self-collaboration\textsuperscript{\textdagger} & 90.2 & 78.9 \\
\midrule
\methodname & \textbf{93.90} & \textbf{91.23} \\
\bottomrule
\end{tabular}}
\end{table}

\textbf{Refinement-based approaches.} Approaches in this category refine the generated code based on feedback, either from the LLM itself or external sources, such as compilers and interpreters.
\begin{itemize}
\item \textbf{Self-collaboration} \cite{self_collaboration} enables multiple LLM agents to act as distinct roles (i.e., analyst, coder, and tester) within a virtual team. These roles interact and collaborate in a specified manner to address code generation tasks.
\item \textbf{Self-repair} \cite{self_repair} employs a feedback model to generate textual explanations for errors encountered during unit test execution. The code model then uses these explanations to repair the initial code. 
\item \textbf{Self-debugging} \cite{self_debugging} teaches LLMs to perform iterative \textit{rubber duck debugging} by explaining the generated code line-by-line with few-shot demonstrations. 
\item \textbf{INTERVENOR} \cite{intervenor} prompts LLMs to play two distinct roles. The Code Teacher iteratively crafts the interactive chain of repair based on compiler feedback, which guides the Code Learner to generate or repair code.
\item \textbf{Reflexion} \cite{reflexion} uses CoT prompting to generate its own test suites. It then iteratively generates code and verbal self-reflections based on this self-generated test feedback, which guide subsequent implementations.
\item \textbf{LDB} \cite{ldb} segments programs into basic blocks and tracks intermediate variable values during runtime execution for iterative program repair.
\item \textbf{MetaGPT} \cite{metagpt} is a multi-agent system that simulates a complete software company by defining five roles and leveraging human-like standard operating procedures.
\end{itemize}

\paragraph{Implementation details}
\label{sec:details}
We apply both open-source and closed-source LLMs to \methodname, including DeepSeek-Coder-Instruct with 33B parameters \cite{deepseekcoder} and GPT-3.5-Turbo (gpt-3.5-turbo-0613) \cite{chatgpt}, where the maximum context window is 16K tokens for all approaches.
Moreover, the \navigator employs text-embedding-3-large\footnote{\url{https://platform.openai.com/docs/guides/embeddings}} to vectorize the $n=15$ sampled solution plans and divides them into $k=3$ clusters.
Programs taking longer than 3 seconds to execute on any test case are marked as \textit{Time Limit Exceeded}.
We adopt a temperature of $0.8$ to sample diverse solution plans and use greedy decoding by setting the temperature to $0$ in other steps.
For the comparative methods, we reproduce them according to the source code or provided prompts; otherwise, we quote the results directly from their papers \cite{ldb,self_collaboration,metagpt}.
Following the prior works \cite{self_debugging,ldb,self_planning,clarifygpt}, we use greedy decoding in the reproduced methods unless otherwise specified.
The maximum number of iterations $r$ is set to $10$ for \methodname and other iterative refinement-based approaches, i.e., up to $10$ times the code is generated or refined.

To ensure a fair comparison, all approaches are given identical problem descriptions and public test cases for code generation. 
The generated code is then executed and evaluated in a consistent Python 3.9 environment. 
This guarantees that the code generated by different approaches will receive consistent external feedback, thereby enabling unbiased and rigorous comparisons.

\subsection{RQ1: Accuracy Comparison}

The accuracy comparison results are presented in Table~\ref{tab:exp-main}.
Our \methodname achieves the best pass@1 scores across all benchmarks and foundation models.
In comparison to prompting LLMs directly, \methodname shows significant relative improvement of 12.00\% to 162.43\%.
The prompting techniques would accumulate errors in intermediate thoughts and single code generation, causing relatively weak accuracy on code generation.
CoT and SCoT prompting are even worse than direct generation in some settings, which is consistent with the findings of prior works \cite{brainstorm, karecoder}. 
The poor performance of Reflexion is likely due to the model generating incorrect test cases, which leads to self-reflections based on false negative evaluations of the code \cite{reflexion}.
In contrast, other refinement-based approaches using provided public test cases achieve overall accuracy gains, since the reliable test feedback can guide the refinement in more promising directions.
However, they are confined to a single solving path, lacking the flexibility to explore alternative solution plans when stuck. 
To overcome these limitations, \methodname combines the advantages of multi-plan exploration and feedback-driven refinement.

Accuracy discrepancies across benchmarks are worth examining. 
Most approaches perform well on the relatively simple HumanEval and MBPP benchmarks. 
However, all approaches exhibit a substantial accuracy decrease on the challenging CodeContest benchmark. This reflects that current code generation techniques still have room for improvement in tackling complex programming problems.
Note that the direct prompting with DeepSeek-Coder even outperforms that with GPT-3.5-Turbo on HumanEval, HumanEval+, and CodeContest-test. 
We speculate this may be due to data leakage issues, which will be further analyzed in Sect.~\ref{sec:threats}.

We further evaluate \methodname with one of the most advanced LLMs, GPT-4 (gpt-4-0613) \cite{gpt4_report}, and cite the results of several powerful approaches \cite{metagpt, reflexion, self_collaboration} from their original papers.
The comparison results are shown in Table~\ref{tab:exp-gpt4}. 
\methodname still significantly improves accuracy over direct prompting and outperforms existing multi-agent approaches.

\begin{tcolorbox}
\textbf{Answer to RQ1:} \methodname outperforms all baselines across five benchmarks with two advanced LLMs. 
Compared with direct generation, it gains remarkable relative improvements ranging from 16.97\% to 162.43\% on GPT-3.5-Turbo, and from 12.00\% to 128.76\% on DeepSeek-Coder.
The largest improvement is achieved on the most challenging CodeContest benchmark.
\end{tcolorbox}


\subsection{RQ2: Hyperparameter Impact}

We investigate the impact of two critical hyperparameters: the maximum number of iterations $r$ and the number of clusters $k$.

\paragraph{Iteration count impact}
Given $1 \leq r \leq 10$, we compare \methodname with two iterative refinement-based approaches, INTERVENOR and Self-debugging.
We report the experimental results on HumanEval and the test set of CodeContest with GPT-3.5-Turbo.

\begin{figure}
\centering
\includegraphics[width=\columnwidth]{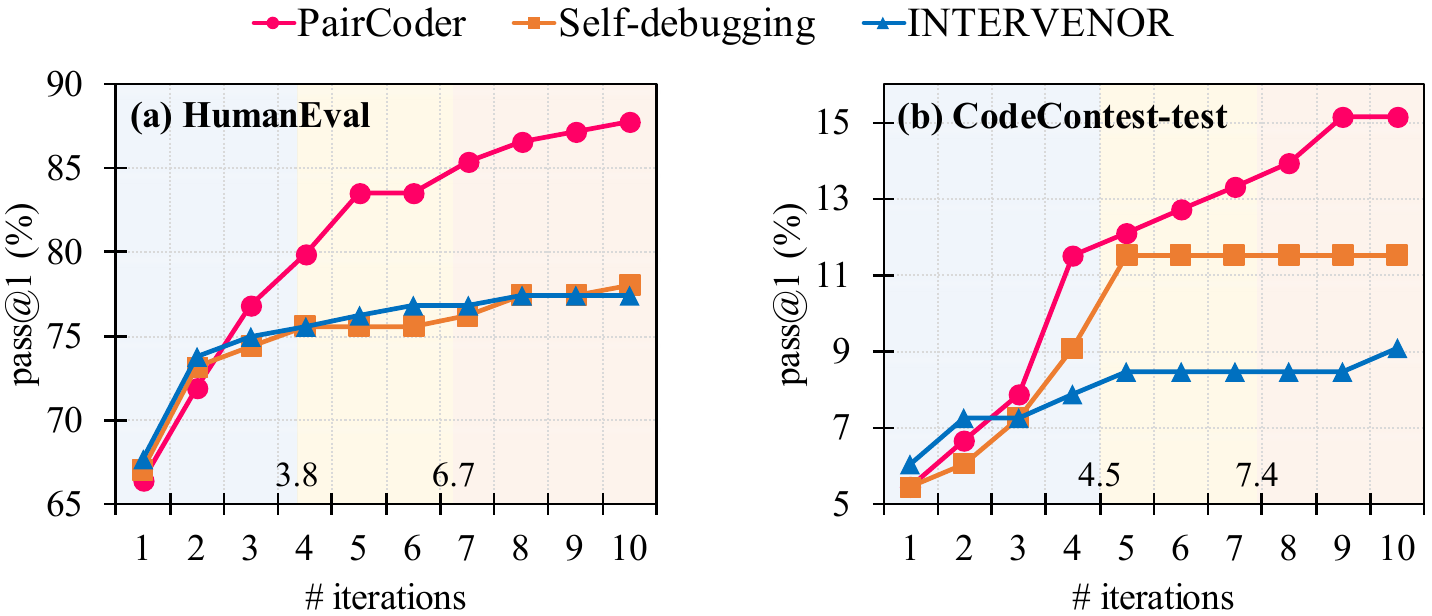}
\caption{Accuracy changes with the maximum number of iterations, which evaluates \methodname, Self-debugging, and INTERVENOR using GPT-3.5-Turbo on (a) HumanEval and (b) CodeContest-test benchmarks. 
The colored demarcations denote the average iteration counts at which \methodname transitions to its next plan.}
\Description{Accuracy changes with the maximum number of iterations, which evaluates \methodname, Self-debugging, and INTERVENOR using GPT-3.5-Turbo on (a) HumanEval and (b) CodeContest-test benchmarks. 
The colored demarcations denote the average iteration counts at which \methodname transitions to its next plan.}
\label{fig:iteration}
\end{figure}

Fig.~\ref{fig:iteration} depicts the line plots that show the variation in the pass@1 metric with increasing iteration count for different approaches.
As the iteration count increases, all approaches exhibit accuracy improvements, but the extent of these improvements varies.
On HumanEval, while Self-debugging and INTERVENOR show modest gains of around 10 percentage points in pass@1 scores after 10 iterations, \methodname shows a substantially larger improvement of over 21 points. A similar pattern emerges on CodeContest-test, where \methodname's pass@1 score raises by nearly 10 points, outpacing the limited improvements of the two baselines.

Notably, Self-debugging and INTERVENOR appear to reach an accuracy plateau after a certain number of iterations, e.g., $r \ge 5$ on CodeContest-test. 
This observation aligns with prior findings \cite{ldb, self_debugging}. 
In contrast, \methodname consistently maintains an upward trajectory across all iteration counts on both benchmarks, suggesting its ability to leverage more iterations for continuous accuracy improvement.
Furthermore, we examine when \methodname typically switches to a new solution plan during iterative refinement. Fig.~\ref{fig:iteration} also labels the average iteration counts at which \methodname transitions to its next plan, denoted by the colored demarcations, e.g., 3.8 and 6.7 on HumanEval.
They also reflect the average count of code repairs under each plan.
We observe that the demarcations roughly align with the iteration counts where Self-debugging and INTERVENOR appear to plateau.
This correlation suggests that multi-plan exploration based on the long-term memory module enables \methodname to effectively identify and abandon unpromising plans.
Within the same maximum number of iterations, \methodname can timely explore multiple candidate plans based on execution feedback, achieving superior accuracy by expanding the search space away from local optima.

\begin{figure}
\centering
\includegraphics[width=\columnwidth]{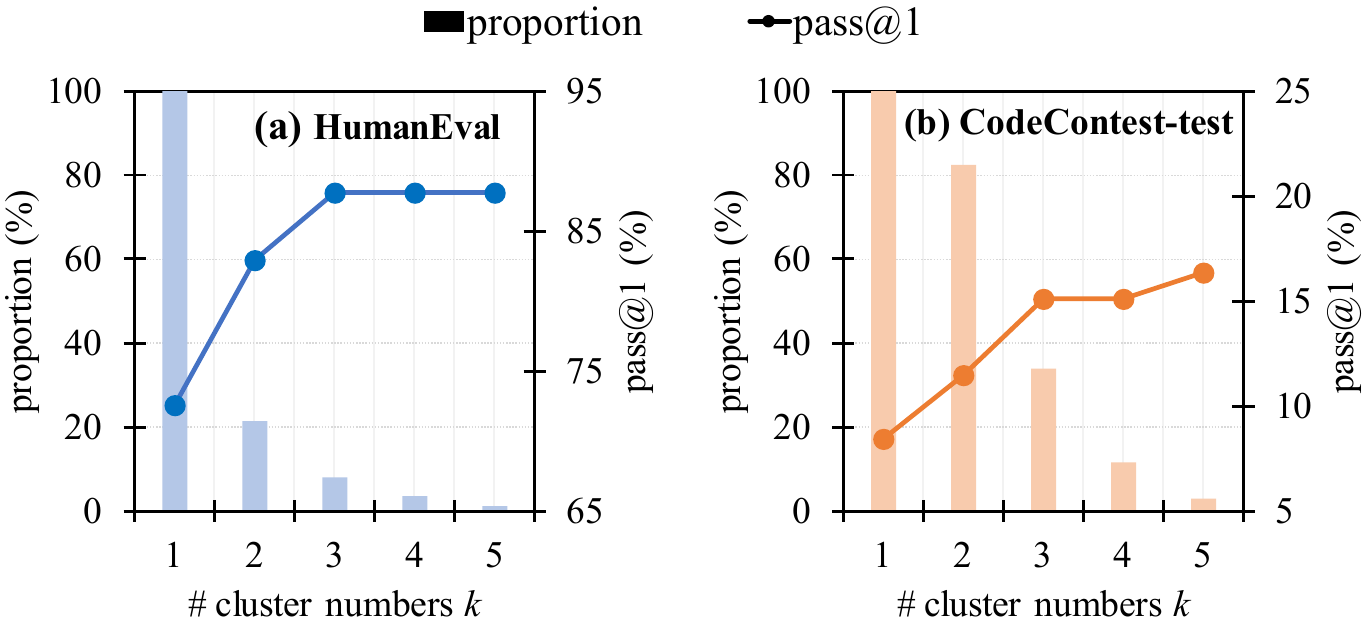}
\caption{Accuracy changes with the number of clusters, which evaluates \methodname by pass@1 score and the proportion of problems where all $k$ plans are attempted.}
\Description{Accuracy changes with the number of clusters, which evaluates \methodname by pass@1 score and the proportion of problems where all $k$ plans are attempted.}
\label{fig:cluster}
\end{figure}

\paragraph{Cluster number impact}
To investigate the impact of the cluster number $k$ on \methodname, we vary $k$ from 1 to 5.
We also report the results on HumanEval and CodeContest-test with GPT-3.5-Turbo.

As shown in Fig.~\ref{fig:cluster}, the line plots reflect the changes in pass@1, and the bar charts are for the proportion of problems where all $k$ plans are attempted.
As the cluster number $k$ increases, the pass@1 score of \methodname improves but reaches a bottleneck after $k=3$.
Intuitively, a larger $k$ can bring more candidate solution plans for iterative refinement, leading to a positive correlation with the accuracy of \methodname.
However, only 4.88\% and 14.55\% of the problems in HumanEval and CodeContest-test, respectively, attempt more than three plans. 
This is because the iterative process would stop early due to passing public test cases or reaching the maximum number of iterations, resulting in a plateau in accuracy.
It suggests that a moderate cluster number $k=3$ appears to be optimal for \methodname with the maximum number of iterations $r=10$. 
We think that a larger iteration count and cluster number may bring more improvement for \methodname, but it is a trade-off between effectiveness and cost.

The proportions of attempting all $k$ plans differ significantly between the two benchmarks. 
For the simpler HumanEval, only 21.34\% of problems attempt the second plan with $k=2$, while this value is 82.42\% for CodeContest-test.
CodeContest consists of competition-level programming problems, where an initial solution plan is likely to fail to solve the problem.
This observation on complex problems supports our design for multi-plan exploration.

\begin{tcolorbox}
\textbf{Answer to RQ2:} The iteration count and cluster number significantly affect the accuracy of \methodname. 
The increase in iteration count brings a substantial and consistent improvement to \methodname versus the baselines.
With a maximum of 10 iteration rounds, the optimal cluster number is $3$, which is a trade-off for cost efficiency. 
\end{tcolorbox}

\subsection{RQ3: Ablation Study}

To analyze the individual effectiveness of multi-plan exploration and feedback-driven refinement in \methodname, we conduct ablation studies in Table~\ref{tab:exp-abl}.
``w/o MP'' disables the capability of multi-plan exploration, making the \navigator always choose to repair the current code rather than adjust the solution plan in Step 5.
``w/o RF'' is the opposite, which disables the feedback-driven refinement process, making the \navigator always choose to attempt another candidate plan.
The \driver's behavior changes according to different directions of the \navigator.

The ablation results demonstrate that the complete \methodname achieves the best accuracy, and both multi-plan exploration and feedback-driven refinement play a positive role in code generation.
Multi-plan exploration effectively expands the search space away from local optima, leading to considerable improvements.
Feedback-driven refinement brings more significant improvement than multi-plan exploration across all benchmarks and foundation models.
It indicates that even with a proper solution plan, advanced LLMs still struggle to generate the correct code in one attempt.
The two components are more effective for complex problems, yielding substantial relative improvements of 38.86\% and 56.35\% using GPT-3.5-Turbo on CodeContest-test, respectively, compared to 8.26\% and 18.03\% on the simpler HumanEval.
This is in line with real-world programming practices, where complex problem-solving requires more exploration and refinement.

\begin{table}
\caption{Ablation results for multi-plan exploration (MP) and feedback-driven refinement (RF).}
\label{tab:exp-abl}
\centering
\resizebox{\linewidth}{!}{
\begin{tabular}{c|l|cccccc}
\toprule
\multirow{2}{*}{Models} & \multirow{2}{*}{Variants} & \multicolumn{2}{c}{HumanEval} & \multicolumn{2}{c}{MBPP} & \multicolumn{2}{c}{CodeContest} \\
\cmidrule(lr){3-4} \cmidrule(lr){5-6} \cmidrule(lr){7-8}
& & Orig & Plus & Orig & Plus & Valid & Test \\
\midrule
\multirow{3}{*}{\makecell{GPT-3.5\\-Turbo}} 
& \methodname    & \textbf{87.80} & \textbf{77.44} & \textbf{80.60} & \textbf{77.69} & \textbf{17.95} & \textbf{15.15} \\
& \quad w/o MP   & 81.10 & 73.78 & 76.20 & 71.68 &  11.97 & 10.91 \\
& \quad w/o RF   & 74.39 & 68.29 & 72.20 & 68.92 & \ \ 8.55 & \ \ 9.69 \\
\midrule
\multirow{3}{*}{\makecell{DeepSeek\\-Coder}} 
& \methodname    & \textbf{85.37} & \textbf{76.22} & \textbf{78.80} & \textbf{75.69} & \textbf{13.68} & \textbf{14.55} \\
& \quad w/o MP   & 78.86 & 73.17 & 73.40 & 71.93 & \ \ 9.40 & 12.12 \\
& \quad w/o RF   & 75.61 & 69.51 & 68.60 & 67.92 & \ \ 7.69 & 10.91 \\
\bottomrule
\end{tabular}
}
\end{table}

\begin{tcolorbox}
\textbf{Answer to RQ3:} Both multi-plan exploration and feedback-driven refinement contribute to \methodname for code generation. 
They bring more significant improvements for complex problems in CodeContest, which is in line with real-world programming practices.
\end{tcolorbox}

\subsection{RQ4: Cost and Error Analyses}

For this RQ, we further investigate the usage of \methodname.

\paragraph{Cost analysis}
We perform a cost analysis for \methodname and all reproduced approaches on HumanEval, MBPP, and the test set of CodeContest.
The cost is measured by two key metrics: the average number of API calls per problem and the average token consumption per problem.
For each approach, we record its API requests and responses using GPT-3.5-Turbo.
This provides the number of API calls and token consumption, including input tokens and generated output tokens of LLMs.
Note that we count API calls to assess the efficiency in code generation, since the time spent is susceptible to uncontrollable factors such as network fluctuation.

For the fairness of comparison, we extend the comparative methods to conduct additional experiments using GPT-3.5-turbo: (i) For prompting techniques, we repeat sampling with a temperature of $0.8$ until the generated code passes all public test cases or $10$ attempts are reached. (ii) For Self-repair, we also allow up to $10$ iterations. This setup ensures that these approaches have the same maximum number of attempts as the iterative approaches in Sect.~\ref{sec:details}.
As shown in Table~\ref{tab:exp-repeat}, the results demonstrate that simple repetitive sampling can indeed enhance these approaches, but \methodname remains dominant. 
It confirms that the effectiveness of \methodname beyond merely increased computation. 

\begin{table}[!t]
\centering
\caption{Accuracy comparison of prompting approaches and Self-repair using GPT-3.5-Turbo with up to 10 iterations. HumanEval and MBPP are the original versions (the same below unless otherwise specified).}
\label{tab:exp-repeat}
{\small
\begin{tabular}{l|ccc}
\toprule
Approaches & HumanEval & MBPP & CodeContest-test \\
\midrule
Direct prompting & 75.61 & 72.80 & 9.09 \\
CoT prompting & 76.83 & 73.00 & 7.88 \\
SCoT prompting & 75.61 & 73.94 & 8.48 \\
Self-planning & 79.27 & 76.36 & 10.90 \\
Self-repair & 78.65 & 75.00 & 9.09 \\
\midrule
\methodname & \textbf{87.80} & \textbf{80.60} & \textbf{15.15}\\
\bottomrule
\end{tabular}}
\end{table}

\begin{table}
\centering
\caption{Average number of API calls and tokens (in thousands) using GPT-3.5-Turbo. Note that all approaches are allowed up to 10 iterations.}
\label{tab:exp-cost}
{\small
\begin{tabular}{l|cc|cc|cc}
\toprule
\multirow{2}{*}{Approaches} & \multicolumn{2}{c|}{HumanEval} & \multicolumn{2}{c|}{MBPP} & \multicolumn{2}{c}{CodeContest-test} \\
\cmidrule(lr){2-3}\cmidrule(lr){4-5}\cmidrule(lr){6-7}
& API & Token & API & Token & API & Token  \\
\midrule
\multirow{1}{*}{Direct prompting}  
& 2.43 & 1.19 & 2.31 & 1.23 & \ \ 8.73 & \ \ 9.79 \\
\multirow{1}{*}{CoT prompting} 
& 2.43 & 1.36 & 2.37 & 1.41 & \ \ 8.62 & \ \ 10.16 \\
\multirow{1}{*}{SCoT prompting} 
& 4.63 & 4.46 & 4.61 & 4.49 & \ \ 17.30 & \ \ 26.57 \\
\multirow{1}{*}{Self-planning} 
& 4.16 & 3.27 & 4.67 & 2.78 & \ \ 17.02 & \ \ 20.03 \\
\midrule
\multirow{1}{*}{Self-repair}  
& 2.37 & 2.40 & 2.36 & 2.02 & \ \ 8.85 & \ \ 19.82 \\
\multirow{1}{*}{Reflexion}  
& \textbf{8.47} & 9.68 & \textbf{9.58} & \textbf{10.28} & \ \ - & \ \ - \\
\multirow{1}{*}{LDB\textsuperscript{\textdagger}}  
& - & \textbf{23} & - & - & \ \ - & \ \ - \\
\multirow{1}{*}{Self-debugging}   
& 2.74 & 7.74 & 2.55 & 6.47 & \ \ 9.07 & \textbf{38.40} \\
\multirow{1}{*}{INTERVENOR}  
& 3.85 & 2.05 & 3.77 & 1.83 & 17.06 & 13.72 \\
\midrule
\multirow{1}{*}{\methodname} 
& 5.28 & 5.98 & 5.37 & 5.09 & \textbf{17.56} & 24.06 \\
\bottomrule
\end{tabular}}
\end{table}

Table~\ref{tab:exp-cost} presents the cost comparison results.
Among prompting techniques, Direct and CoT prompting generally have lower costs compared to other approaches, while SCoT and Self-planning demonstrate higher costs due to their complex prompting strategies. 
For refinement-based approached, Reflexion shows high API call and token consumption on both benchmarks. LDB, reported only for HumanEval, demonstrates the highest token consumption among all approaches.
Collaborating the \navigator and the \driver, \methodname requires more API calls than most approaches except Reflexion. 
Nevertheless, \methodname maintains moderate token consumption, particularly when compared to LDB, Reflexion, and Self-debugging across different benchmarks.
Furthermore, we observe that all approaches spend higher cost on the challenging CodeContest-test than simpler benchmarks, due to the fact that complex problem-solving requires more model interactions.
Despite moderate cost increase, \methodname significantly improves accuracy across all settings (as shown in Tables~\ref{tab:exp-main} and ~\ref{tab:exp-repeat}), justifying the minor increase cost. 
Overall, \methodname achieves better accuracy in code generation while maintaining reasonable costs compared to existing approaches.

\paragraph{Error analysis}
We conduct a detailed analysis of the error types encountered by \methodname using GPT-3.5-Turbo on three benchmarks. Table~\ref{tab:error_analysis} presents the analysis results.
Overall, \textit{Wrong Answer} is the most common error type, indicating that generating functionally correct programs remains a key challenge for code generation.
On the relatively simple HumanEval and MBPP benchmarks, \methodname solves over 80\% programming problems, where \textit{Runtime Error} and \textit{Time Limit Exceeded} are rare.
However, on the challenging CodeContest-test benchmark, the accuracy largely decreases. Although the causes of errors are still dominated by \textit{Wrong Answer}, there is a notable increase in \textit{Runtime Error} and \textit{Time Limit Exceeded}. 
The analysis results are consistent with realistic programming practices, where simple problems generally do not encounter unexpected exceptions and efficiency issues.
We emphasize the urgent need to further improve the functional correctness of code generation, and also focus on the efficiency and robustness for complex programming problems.

Since only the public test cases $T_v$ are visible during the code generation and refinement stages, their quality directly impacts the final performance.
On HumanEval and MBPP, the pass rates on public test cases are over 93\%, and 91.72\% (87.80/95.73) and 86.48\% (80.60/93.20) of them also pass the private test cases $T_h$, respectively.
However, on the challenging CodeContest-test, the pass rate on $T_v$ is only 21.21\%, and 71.43\% (15.15/21.21) of them eventually pass $T_h$.
As with code testing, public test cases can provide real feedback from the executor, revealing the vulnerability of a program to specific inputs.
Referring to Table~\ref{tab:benchmark_statistics}, the low coverage of public test cases limits the ability of \methodname to facilitate code generation.

Based on the above findings, the evolution of code generation approaches and the expansion of public test cases are both crucial and orthogonal.
\methodname seems powerful enough for simple programming problems, and broader public test cases such as edge cases would bring the accuracy closer to that on the current $T_v$.
For complex problems, it is imperative to enhance the reasoning and programming capabilities inherent in LLMs.
Besides, retrieval augmentation \cite{Ren2023,karecoder,Su2024ARKS} and test case generation \cite{codet,zhang2023algo,ridnik2024code,reflexion} are potential future improvements in code generation.

\begin{tcolorbox}
\textbf{Answer to RQ4:} 
Compared to existing approaches, \methodname achieves superior accuracy with comparable and reasonable cost.
\textit{Wrong Answer} is the main error cause, yet efficiency and robustness also deserve concerns for complex problems.
In addition to enhancing code generation, test case generation is a promising orthogonal direction.
\end{tcolorbox}

\begin{table}
\centering
\caption{Analysis of passes on public test cases $T_v$, private test cases $T_h$, and specific error types on $T_h$, i.e., \textit{Runtime Error} (RE), \textit{Wrong Answer} (WA), and \textit{Time Limit Exceeded} (TLE).}
\label{tab:error_analysis}
{\small
\begin{tabular}{l|c|c|ccc}
\toprule
\multirow{2}{*}{Benchmarks} & \multirow{2}{*}{Pass $T_v$} & \multirow{2}{*}{Pass $T_h$} & \multicolumn{3}{c}{Not pass $T_h$} \\
\cmidrule(lr){4-6}
& & & RE & WA & TLE \\
\midrule
HumanEval & 95.73\% & 87.80\% & \ \ 0 & 11.59\% & 0.61\% \\
MBPP & 93.20\% & 80.60\% & \ \ 0.80\% & 18.60\% & 0 \\
CodeContest-test & 21.21\% & 15.15\% & 16.36\% & 60.00\% & 8.48\% \\
\bottomrule
\end{tabular}}
\end{table}

\subsection{Threats to Validity}
\label{sec:threats}

The first potential threat relates to the generalizability of our evaluation. 
To mitigate this concern, we carefully select five widely used and representative benchmarks and both closed-source and open-source LLMs for our experiments. 
Under all these settings, consistently superior accuracy demonstrates the effectiveness of \methodname. 
In future work, we plan to further validate the generalizability of \methodname across a broader range of LLMs and benchmarks, such as multilingual code generation \cite{Athiwaratkun2023Multi, Zheng2023CodeGeeX}.

Another potential threat is data leakage in pre-trained LLMs. 
For example, DeepSeek-Coder was released after benchmarks like HumanEval were collected, raising the possibility that benchmark samples were unintentionally included in its pre-training corpus. 
However, any such leakage would affect all baselines equally since they use the same model. 
Therefore, while leakage may inflate overall accuracy, it does not affect the fairness of our comparative analysis and the relative gains of \methodname, which consistently shows the largest improvements across all settings.

The third potential threat arises from the zero-shot prompting used in our experiments. 
Although zero-shot prompting achieves superior accuracy while largely reducing token consumption, we do not rule out the possibility that other instructions or few-shot demonstrations could further improve performance. 
However, the selection of demonstrations for in-context learning poses a significant challenge, which can greatly influence the behavior of LLMs \cite{Gao_2023}. 
We leave the exploration of effective few-shot prompting techniques in future work.


\section{Related Work}
\label{sec:work}
Recent advancements in LLMs have shown remarkable capabilities in code generation tasks by training on vast amounts of code-containing corpora. 
Open-source models like InCoder \cite{incoder}, Code Llama \cite{codellama}, WizardCoder \cite{wizardcoder}, and DeepSeek-Coder \cite{deepseekcoder}, have depicted performance that matches or even surpasses closed-source commercial models like ChatGPT \cite{chatgpt}, GPT-4 \cite{gpt4_report}, and Claude~\cite{claude}.
This development has significantly improved software productivity \cite{Kazemitabaar2023Studying,Peng2023The} and profoundly affected the progress of intelligent software engineering, attracting substantial work focused on enhancing the code generation capabilities of LLMs.

A key focus is on prompting techniques that guide LLMs to produce intermediate reasoning steps from problem descriptions. 
Prompting techniques have been proven to effectively improve the code generation performance of LLMs in a plug-and-play manner \cite{cot, scot, self_planning, brainstorm, codecot}. 
SCoT \cite{scot} and Self-planning \cite{self_planning} design different formats of intermediate steps, while BRAINSTORM \cite{brainstorm} trains a neural ranker model to select the best thought. 
They leverage prompting techniques to stimulate the reasoning capabilities of LLMs, guiding them to generate more accurate code.
However, generating correct code is rarely a one-time effort \cite{self_debugging}. 
Some approaches \cite{codet, alphacode, lever, mbr_exe, codereviewer, coderanker} first generate multiple code solutions and then filter or rank them based on consistency or execution results to obtain the final code. 
They require substantial computational resources to generate code candidates, which is inefficient and orthogonal to our framework.

Another line of work focuses on refining the generated code \cite{self_edit, self_repair, self_debugging, intervenor, ldb}, where the feedback is obtained from LLMs themselves or external sources. 
Self-edit \cite{self_edit} trains a fault-aware code editor that employs error messages to refine the generated code. 
Self-repair \cite{self_repair} investigates the effects of leveraging feedback from diverse sources for code repair, such as humans or LLMs.
Self-debugging \cite{self_debugging} utilizes explanations generated by LLMs to repair code. 
INTERVENOR \cite{intervenor} emulates the interactive code repair processes based on compiler feedback. 
LDB \cite{ldb} mainly focuses on debugging by leveraging runtime execution information, which is orthogonal to our work. It can be integrated into our \methodname framework to enhance code repair in Steps 5 and 6.
We include competitive baselines in our experiments \cite{scot,self_planning,self_repair,self_debugging,intervenor, ldb}, as shown in Table~\ref{tab:exp-main},  exclude Self-Edit \cite{self_edit} due to its additional training requirements.

Additionally, several works \cite{self_collaboration, chatdev, metagpt} employ collaborative LLM agents to simulate the full software development lifecycle such as the waterfall model \cite{waterfull}, spanning from high-level tasks like requirements analysis and architecture design \cite{chatdev, metagpt}, to low-level roles like coder and tester \cite{self_collaboration}. A notable example is MetaGPT \cite{metagpt}. 
In contrast, our \methodname focuses specifically on the critical coding stage, essentially corresponding to the Engineer agent in MetaGPT. 
Instead of attempting to simulate the entire development lifecycle, \methodname uses collaborative agents to emulate pair programming, a well-established and proven software practice directly targeting efficient and high-quality code generation. 
While differing in scope, our work complements those broader development lifecycle simulations by concentrating on the essential coding component.

\section{Conclusion}
\label{sec:concl}

In this paper, we propose the \methodname framework, which is the first to adapt pair programming practices into LLM-based code generation. 
It comprises a \navigator agent for high-level planning and a \driver agent for specific implementation, collaborating on code generation via multi-plan exploration and feedback-driven refinement.
The \navigator explores multiple plans based on execution feedback from the \driver and historical memory.
The \driver follows the guidance of the \navigator to undertake initial code generation, code testing, and refinement.
Extensive experiments on diverse benchmarks and LLMs demonstrate the superior accuracy of \methodname. 
Our work represents a promising step towards leveraging collaborative agents to facilitate intelligent software development.
In future work, we plan to integrate human feedback or external knowledge sources to further enhance the high-level planning capabilities of the \navigator.
We will also explore applications of the \methodname framework to other domains beyond code generation.

\begin{acks}
We thank the anonymous reviewers for their valuable comments. 
This work was supported by the National Natural Science Foundation of China (No. 62272219) and the Collaborative Innovation Center of Novel Software Technology \& Industrialization.
\end{acks}


\bibliographystyle{ACM-Reference-Format}
\bibliography{ref}


\begin{thebibliography}{54}


\ifx \showCODEN    \undefined \def \showCODEN     #1{\unskip}     \fi
\ifx \showDOI      \undefined \def \showDOI       #1{#1}\fi
\ifx \showISBNx    \undefined \def \showISBNx     #1{\unskip}     \fi
\ifx \showISBNxiii \undefined \def \showISBNxiii  #1{\unskip}     \fi
\ifx \showISSN     \undefined \def \showISSN      #1{\unskip}     \fi
\ifx \showLCCN     \undefined \def \showLCCN      #1{\unskip}     \fi
\ifx \shownote     \undefined \def \shownote      #1{#1}          \fi
\ifx \showarticletitle \undefined \def \showarticletitle #1{#1}   \fi
\ifx \showURL      \undefined \def \showURL       {\relax}        \fi
\providecommand\bibfield[2]{#2}
\providecommand\bibinfo[2]{#2}
\providecommand\natexlab[1]{#1}
\providecommand\showeprint[2][]{arXiv:#2}

\bibitem[Anthropic(2023)]%
        {claude}
\bibfield{author}{\bibinfo{person}{Anthropic}.} \bibinfo{year}{2023}\natexlab{}.
\newblock \bibinfo{title}{Meet {Claude}}.
\newblock \bibinfo{howpublished}{\url{https://www.anthropic.com/claude/}}.
\newblock
\newblock
\shownote{Accessed May 30, 2024}.


\bibitem[Arthur and Vassilvitskii(2007)]%
        {kmeans}
\bibfield{author}{\bibinfo{person}{David Arthur} {and} \bibinfo{person}{Sergei Vassilvitskii}.} \bibinfo{year}{2007}\natexlab{}.
\newblock \showarticletitle{k-means++: The Advantages of Careful Seeding}. In \bibinfo{booktitle}{\emph{SODA}}. \bibinfo{publisher}{SIAM}, \bibinfo{address}{New Orleans, LA, USA}, \bibinfo{pages}{1027–1035}.
\newblock


\bibitem[Athiwaratkun et~al\mbox{.}(2023)]%
        {Athiwaratkun2023Multi}
\bibfield{author}{\bibinfo{person}{Ben Athiwaratkun}, \bibinfo{person}{Sanjay~Krishna Gouda}, \bibinfo{person}{Zijian Wang}, \bibinfo{person}{Xiaopeng Li}, \bibinfo{person}{Yuchen Tian}, \bibinfo{person}{Ming Tan}, \bibinfo{person}{Wasi~Uddin Ahmad}, \bibinfo{person}{Shiqi Wang}, \bibinfo{person}{Qing Sun}, \bibinfo{person}{Mingyue Shang}, \bibinfo{person}{Sujan~Kumar Gonugondla}, \bibinfo{person}{Hantian Ding}, \bibinfo{person}{Varun Kumar}, \bibinfo{person}{Nathan Fulton}, \bibinfo{person}{Arash Farahani}, \bibinfo{person}{Siddhartha Jain}, \bibinfo{person}{Robert Giaquinto}, \bibinfo{person}{Haifeng Qian}, \bibinfo{person}{Murali~Krishna Ramanathan}, {and} \bibinfo{person}{Ramesh Nallapati}.} \bibinfo{year}{2023}\natexlab{}.
\newblock \showarticletitle{Multi-lingual Evaluation of Code Generation Models}. In \bibinfo{booktitle}{\emph{ICLR}}. \bibinfo{publisher}{OpenReview.net}, \bibinfo{address}{Kigali, Rwanda}, \bibinfo{pages}{1--19}.
\newblock


\bibitem[Austin et~al\mbox{.}(2021)]%
        {mbpp}
\bibfield{author}{\bibinfo{person}{Jacob Austin}, \bibinfo{person}{Augustus Odena}, \bibinfo{person}{Maxwell~I. Nye}, \bibinfo{person}{Maarten Bosma}, \bibinfo{person}{Henryk Michalewski}, \bibinfo{person}{David Dohan}, \bibinfo{person}{Ellen Jiang}, \bibinfo{person}{Carrie~J. Cai}, \bibinfo{person}{Michael Terry}, \bibinfo{person}{Quoc~V. Le}, {and} \bibinfo{person}{Charles Sutton}.} \bibinfo{year}{2021}\natexlab{}.
\newblock \showarticletitle{Program Synthesis with Large Language Models}.
\newblock \bibinfo{journal}{\emph{CoRR}}  \bibinfo{volume}{2108.07732} (\bibinfo{year}{2021}), \bibinfo{pages}{1--34}.
\newblock


\bibitem[Chen et~al\mbox{.}(2023)]%
        {codet}
\bibfield{author}{\bibinfo{person}{Bei Chen}, \bibinfo{person}{Fengji Zhang}, \bibinfo{person}{Anh Nguyen}, \bibinfo{person}{Daoguang Zan}, \bibinfo{person}{Zeqi Lin}, \bibinfo{person}{Jian{-}Guang Lou}, {and} \bibinfo{person}{Weizhu Chen}.} \bibinfo{year}{2023}\natexlab{}.
\newblock \showarticletitle{CodeT: Code Generation with Generated Tests}. In \bibinfo{booktitle}{\emph{ICLR}}. \bibinfo{publisher}{OpenReview.net}, \bibinfo{address}{Kigali, Rwanda}, \bibinfo{pages}{1--19}.
\newblock


\bibitem[Chen et~al\mbox{.}(2021)]%
        {humaneval}
\bibfield{author}{\bibinfo{person}{Mark Chen}, \bibinfo{person}{Jerry Tworek}, \bibinfo{person}{Heewoo Jun}, \bibinfo{person}{Qiming Yuan}, \bibinfo{person}{Henrique~Pond{\'{e}} de Oliveira~Pinto}, \bibinfo{person}{Jared Kaplan}, \bibinfo{person}{Harrison Edwards}, \bibinfo{person}{Yuri Burda}, \bibinfo{person}{Nicholas Joseph}, \bibinfo{person}{Greg Brockman}, \bibinfo{person}{Alex Ray}, \bibinfo{person}{Raul Puri}, \bibinfo{person}{Gretchen Krueger}, \bibinfo{person}{Michael Petrov}, \bibinfo{person}{Heidy Khlaaf}, \bibinfo{person}{Girish Sastry}, \bibinfo{person}{Pamela Mishkin}, \bibinfo{person}{Brooke Chan}, \bibinfo{person}{Scott Gray}, \bibinfo{person}{Nick Ryder}, \bibinfo{person}{Mikhail Pavlov}, \bibinfo{person}{Alethea Power}, \bibinfo{person}{Lukasz Kaiser}, \bibinfo{person}{Mohammad Bavarian}, \bibinfo{person}{Clemens Winter}, \bibinfo{person}{Philippe Tillet}, \bibinfo{person}{Felipe~Petroski Such}, \bibinfo{person}{Dave Cummings}, \bibinfo{person}{Matthias Plappert}, \bibinfo{person}{Fotios
  Chantzis}, \bibinfo{person}{Elizabeth Barnes}, \bibinfo{person}{Ariel Herbert{-}Voss}, \bibinfo{person}{William~Hebgen Guss}, \bibinfo{person}{Alex Nichol}, \bibinfo{person}{Alex Paino}, \bibinfo{person}{Nikolas Tezak}, \bibinfo{person}{Jie Tang}, \bibinfo{person}{Igor Babuschkin}, \bibinfo{person}{Suchir Balaji}, \bibinfo{person}{Shantanu Jain}, \bibinfo{person}{William Saunders}, \bibinfo{person}{Christopher Hesse}, \bibinfo{person}{Andrew~N. Carr}, \bibinfo{person}{Jan Leike}, \bibinfo{person}{Joshua Achiam}, \bibinfo{person}{Vedant Misra}, \bibinfo{person}{Evan Morikawa}, \bibinfo{person}{Alec Radford}, \bibinfo{person}{Matthew Knight}, \bibinfo{person}{Miles Brundage}, \bibinfo{person}{Mira Murati}, \bibinfo{person}{Katie Mayer}, \bibinfo{person}{Peter Welinder}, \bibinfo{person}{Bob McGrew}, \bibinfo{person}{Dario Amodei}, \bibinfo{person}{Sam McCandlish}, \bibinfo{person}{Ilya Sutskever}, {and} \bibinfo{person}{Wojciech Zaremba}.} \bibinfo{year}{2021}\natexlab{}.
\newblock \showarticletitle{Evaluating Large Language Models Trained on Code}.
\newblock \bibinfo{journal}{\emph{CoRR}}  \bibinfo{volume}{2107.03374} (\bibinfo{year}{2021}), \bibinfo{pages}{1--35}.
\newblock


\bibitem[Chen et~al\mbox{.}(2024)]%
        {self_debugging}
\bibfield{author}{\bibinfo{person}{Xinyun Chen}, \bibinfo{person}{Maxwell Lin}, \bibinfo{person}{Nathanael Sch{\"a}rli}, {and} \bibinfo{person}{Denny Zhou}.} \bibinfo{year}{2024}\natexlab{}.
\newblock \showarticletitle{Teaching Large Language Models to Self-Debug}. In \bibinfo{booktitle}{\emph{ICLR}}. \bibinfo{publisher}{OpenReview.net}, \bibinfo{address}{Vienna, Austria}, \bibinfo{pages}{1--80}.
\newblock


\bibitem[Cheng et~al\mbox{.}(2023)]%
        {batch_prompting}
\bibfield{author}{\bibinfo{person}{Zhoujun Cheng}, \bibinfo{person}{Jungo Kasai}, {and} \bibinfo{person}{Tao Yu}.} \bibinfo{year}{2023}\natexlab{}.
\newblock \showarticletitle{Batch Prompting: Efficient Inference with Large Language Model APIs}. In \bibinfo{booktitle}{\emph{EMNLP}}. \bibinfo{publisher}{ACL}, \bibinfo{address}{Singapore}, \bibinfo{pages}{792--810}.
\newblock


\bibitem[Deng et~al\mbox{.}(2024)]%
        {Deng2024Large}
\bibfield{author}{\bibinfo{person}{Yinlin Deng}, \bibinfo{person}{Chunqiu~Steven Xia}, \bibinfo{person}{Chenyuan Yang}, \bibinfo{person}{Shizhuo~Dylan Zhang}, \bibinfo{person}{Shujing Yang}, {and} \bibinfo{person}{Lingming Zhang}.} \bibinfo{year}{2024}\natexlab{}.
\newblock \showarticletitle{Large Language Models are Edge-Case Generators: Crafting Unusual Programs for Fuzzing Deep Learning Libraries}. In \bibinfo{booktitle}{\emph{ICSE}}. \bibinfo{publisher}{IEEE/ACM}, \bibinfo{address}{Lisbon, Portugal}, \bibinfo{pages}{70:1--70:13}.
\newblock


\bibitem[Dong et~al\mbox{.}(2023)]%
        {self_collaboration}
\bibfield{author}{\bibinfo{person}{Yihong Dong}, \bibinfo{person}{Xue Jiang}, \bibinfo{person}{Zhi Jin}, {and} \bibinfo{person}{Ge Li}.} \bibinfo{year}{2023}\natexlab{}.
\newblock \showarticletitle{Self-collaboration Code Generation via ChatGPT}.
\newblock \bibinfo{journal}{\emph{CoRR}}  \bibinfo{volume}{2304.07590} (\bibinfo{year}{2023}), \bibinfo{pages}{1--38}.
\newblock


\bibitem[Fried et~al\mbox{.}(2023)]%
        {incoder}
\bibfield{author}{\bibinfo{person}{Daniel Fried}, \bibinfo{person}{Armen Aghajanyan}, \bibinfo{person}{Jessy Lin}, \bibinfo{person}{Sida Wang}, \bibinfo{person}{Eric Wallace}, \bibinfo{person}{Freda Shi}, \bibinfo{person}{Ruiqi Zhong}, \bibinfo{person}{Scott Yih}, \bibinfo{person}{Luke Zettlemoyer}, {and} \bibinfo{person}{Mike Lewis}.} \bibinfo{year}{2023}\natexlab{}.
\newblock \showarticletitle{InCoder: {A} Generative Model for Code Infilling and Synthesis}. In \bibinfo{booktitle}{\emph{ICLR}}. \bibinfo{publisher}{OpenReview.net}, \bibinfo{address}{Kigali, Rwanda}, \bibinfo{pages}{1--26}.
\newblock


\bibitem[Gao et~al\mbox{.}(2023)]%
        {Gao_2023}
\bibfield{author}{\bibinfo{person}{Shuzheng Gao}, \bibinfo{person}{Xin{-}Cheng Wen}, \bibinfo{person}{Cuiyun Gao}, \bibinfo{person}{Wenxuan Wang}, \bibinfo{person}{Hongyu Zhang}, {and} \bibinfo{person}{Michael~R. Lyu}.} \bibinfo{year}{2023}\natexlab{}.
\newblock \showarticletitle{What Makes Good In-Context Demonstrations for Code Intelligence Tasks with LLMs?}. In \bibinfo{booktitle}{\emph{ASE}}. \bibinfo{publisher}{IEEE}, \bibinfo{address}{Kirchberg, Luxembourg}, \bibinfo{pages}{761--773}.
\newblock


\bibitem[Geng et~al\mbox{.}(2024)]%
        {Geng2024Large}
\bibfield{author}{\bibinfo{person}{Mingyang Geng}, \bibinfo{person}{Shangwen Wang}, \bibinfo{person}{Dezun Dong}, \bibinfo{person}{Haotian Wang}, \bibinfo{person}{Ge Li}, \bibinfo{person}{Zhi Jin}, \bibinfo{person}{Xiaoguang Mao}, {and} \bibinfo{person}{Xiangke Liao}.} \bibinfo{year}{2024}\natexlab{}.
\newblock \showarticletitle{Large Language Models are Few-Shot Summarizers: Multi-Intent Comment Generation via In-Context Learning}. In \bibinfo{booktitle}{\emph{ICSE}}. \bibinfo{publisher}{IEEE/ACM}, \bibinfo{address}{Lisbon, Portugal}, \bibinfo{pages}{39:1--39:13}.
\newblock


\bibitem[Guo et~al\mbox{.}(2024)]%
        {deepseekcoder}
\bibfield{author}{\bibinfo{person}{Daya Guo}, \bibinfo{person}{Qihao Zhu}, \bibinfo{person}{Dejian Yang}, \bibinfo{person}{Zhenda Xie}, \bibinfo{person}{Kai Dong}, \bibinfo{person}{Wentao Zhang}, \bibinfo{person}{Guanting Chen}, \bibinfo{person}{Xiao Bi}, \bibinfo{person}{Y. Wu}, \bibinfo{person}{Y.~K. Li}, \bibinfo{person}{Fuli Luo}, \bibinfo{person}{Yingfei Xiong}, {and} \bibinfo{person}{Wenfeng Liang}.} \bibinfo{year}{2024}\natexlab{}.
\newblock \showarticletitle{DeepSeek-Coder: When the Large Language Model Meets Programming - The Rise of Code Intelligence}.
\newblock \bibinfo{journal}{\emph{CoRR}}  \bibinfo{volume}{2401.14196} (\bibinfo{year}{2024}), \bibinfo{pages}{1--23}.
\newblock


\bibitem[Holtzman et~al\mbox{.}(2020)]%
        {Holtzman2020The}
\bibfield{author}{\bibinfo{person}{Ari Holtzman}, \bibinfo{person}{Jan Buys}, \bibinfo{person}{Li Du}, \bibinfo{person}{Maxwell Forbes}, {and} \bibinfo{person}{Yejin Choi}.} \bibinfo{year}{2020}\natexlab{}.
\newblock \showarticletitle{The Curious Case of Neural Text Degeneration}. In \bibinfo{booktitle}{\emph{ICLR}}. \bibinfo{publisher}{OpenReview.net}, \bibinfo{address}{Addis Ababa, Ethiopia}, \bibinfo{pages}{1--16}.
\newblock


\bibitem[Hong et~al\mbox{.}(2024)]%
        {metagpt}
\bibfield{author}{\bibinfo{person}{Sirui Hong}, \bibinfo{person}{Mingchen Zhuge}, \bibinfo{person}{Jonathan Chen}, \bibinfo{person}{Xiawu Zheng}, \bibinfo{person}{Yuheng Cheng}, \bibinfo{person}{Ceyao Zhang}, \bibinfo{person}{Jinlin Wang}, \bibinfo{person}{Zili Wang}, \bibinfo{person}{Steven Ka~Shing Yau}, \bibinfo{person}{Zijuan Lin}, \bibinfo{person}{Liyang Zhou}, \bibinfo{person}{Chenyu Ran}, \bibinfo{person}{Lingfeng Xiao}, \bibinfo{person}{Chenglin Wu}, {and} \bibinfo{person}{Jürgen Schmidhuber}.} \bibinfo{year}{2024}\natexlab{}.
\newblock \showarticletitle{MetaGPT: Meta Programming for A Multi-Agent Collaborative Framework}. In \bibinfo{booktitle}{\emph{ICLR}}. \bibinfo{publisher}{OpenReview.net}, \bibinfo{address}{Vienna, Austria}, \bibinfo{pages}{1--26}.
\newblock


\bibitem[Howden(1978)]%
        {Howden1978Theoretical}
\bibfield{author}{\bibinfo{person}{William~E. Howden}.} \bibinfo{year}{1978}\natexlab{}.
\newblock \showarticletitle{Theoretical and Empirical Studies of Program Testing}.
\newblock \bibinfo{journal}{\emph{IEEE Trans. Softw.}}  \bibinfo{volume}{4} (\bibinfo{year}{1978}), \bibinfo{pages}{293--298}.
\newblock


\bibitem[Huang et~al\mbox{.}(2023)]%
        {codecot}
\bibfield{author}{\bibinfo{person}{Dong Huang}, \bibinfo{person}{Qingwen Bu}, \bibinfo{person}{Yuhao Qing}, {and} \bibinfo{person}{Heming Cui}.} \bibinfo{year}{2023}\natexlab{}.
\newblock \showarticletitle{CodeCoT: Tackling Code Syntax Errors in CoT Reasoning for Code Generation}.
\newblock \bibinfo{journal}{\emph{CoRR}}  \bibinfo{volume}{2308.08784} (\bibinfo{year}{2023}), \bibinfo{pages}{1--20}.
\newblock


\bibitem[Huang et~al\mbox{.}(2024a)]%
        {Huang2024Large}
\bibfield{author}{\bibinfo{person}{Jie Huang}, \bibinfo{person}{Xinyun Chen}, \bibinfo{person}{Swaroop Mishra}, \bibinfo{person}{Huaixiu~Steven Zheng}, \bibinfo{person}{Adams~Wei Yu}, \bibinfo{person}{Xinying Song}, {and} \bibinfo{person}{Denny Zhou}.} \bibinfo{year}{2024}\natexlab{a}.
\newblock \showarticletitle{Large Language Models Cannot Self-Correct Reasoning Yet}. In \bibinfo{booktitle}{\emph{ICLR}}. \bibinfo{publisher}{OpenReview.net}, \bibinfo{address}{Kigali, Rwanda}, \bibinfo{pages}{1--17}.
\newblock


\bibitem[Huang et~al\mbox{.}(2024b)]%
        {karecoder}
\bibfield{author}{\bibinfo{person}{Tao Huang}, \bibinfo{person}{Zhihong Sun}, \bibinfo{person}{Zhi Jin}, \bibinfo{person}{Ge Li}, {and} \bibinfo{person}{Chen Lyu}.} \bibinfo{year}{2024}\natexlab{b}.
\newblock \showarticletitle{Knowledge-Aware Code Generation with Large Language Models}. In \bibinfo{booktitle}{\emph{ICPC}}. \bibinfo{publisher}{IEEE/ACM}, \bibinfo{address}{Lisbon, Portugal}, \bibinfo{pages}{52--63}.
\newblock


\bibitem[Inala et~al\mbox{.}(2022)]%
        {coderanker}
\bibfield{author}{\bibinfo{person}{Jeevana~Priya Inala}, \bibinfo{person}{Chenglong Wang}, \bibinfo{person}{Mei Yang}, \bibinfo{person}{Andr{\'{e}}s Codas}, \bibinfo{person}{Mark Encarnaci{\'{o}}n}, \bibinfo{person}{Shuvendu~K. Lahiri}, \bibinfo{person}{Madanlal Musuvathi}, {and} \bibinfo{person}{Jianfeng Gao}.} \bibinfo{year}{2022}\natexlab{}.
\newblock \showarticletitle{Fault-Aware Neural Code Rankers}. In \bibinfo{booktitle}{\emph{NeurIPS}}. \bibinfo{publisher}{Curran Associates, Inc.}, \bibinfo{address}{New Orleans, LA, USA}, \bibinfo{pages}{13419--13432}.
\newblock


\bibitem[Jiang et~al\mbox{.}(2023)]%
        {self_planning}
\bibfield{author}{\bibinfo{person}{Xue Jiang}, \bibinfo{person}{Yihong Dong}, \bibinfo{person}{Lecheng Wang}, \bibinfo{person}{Qiwei Shang}, {and} \bibinfo{person}{Ge Li}.} \bibinfo{year}{2023}\natexlab{}.
\newblock \showarticletitle{Self-planning Code Generation with Large Language Model}.
\newblock \bibinfo{journal}{\emph{CoRR}}  \bibinfo{volume}{2303.06689} (\bibinfo{year}{2023}), \bibinfo{pages}{1--19}.
\newblock


\bibitem[Kazemitabaar et~al\mbox{.}(2023)]%
        {Kazemitabaar2023Studying}
\bibfield{author}{\bibinfo{person}{Majeed Kazemitabaar}, \bibinfo{person}{Justin Chow}, \bibinfo{person}{Carl Ka~To Ma}, \bibinfo{person}{Barbara~J. Ericson}, \bibinfo{person}{David Weintrop}, {and} \bibinfo{person}{Tovi Grossman}.} \bibinfo{year}{2023}\natexlab{}.
\newblock \showarticletitle{Studying the effect of {AI} Code Generators on Supporting Novice Learners in Introductory Programming}. In \bibinfo{booktitle}{\emph{CHI}}. \bibinfo{publisher}{ACM}, \bibinfo{address}{Hamburg, Germany}, \bibinfo{pages}{455:1--455:23}.
\newblock


\bibitem[Kojima et~al\mbox{.}(2022)]%
        {cot_zero}
\bibfield{author}{\bibinfo{person}{Takeshi Kojima}, \bibinfo{person}{Shixiang~Shane Gu}, \bibinfo{person}{Machel Reid}, \bibinfo{person}{Yutaka Matsuo}, {and} \bibinfo{person}{Yusuke Iwasawa}.} \bibinfo{year}{2022}\natexlab{}.
\newblock \showarticletitle{Large Language Models are Zero-Shot Reasoners}. In \bibinfo{booktitle}{\emph{NeurIPS}}. \bibinfo{publisher}{Curran Associates, Inc.}, \bibinfo{address}{New Orleans, LA, USA}, \bibinfo{pages}{22199--22213}.
\newblock


\bibitem[Li et~al\mbox{.}(2023a)]%
        {scot}
\bibfield{author}{\bibinfo{person}{Jia Li}, \bibinfo{person}{Ge Li}, \bibinfo{person}{Yongmin Li}, {and} \bibinfo{person}{Zhi Jin}.} \bibinfo{year}{2023}\natexlab{a}.
\newblock \showarticletitle{Structured Chain-of-Thought Prompting for Code Generation}.
\newblock \bibinfo{journal}{\emph{CoRR}}  \bibinfo{volume}{2305.06599} (\bibinfo{year}{2023}), \bibinfo{pages}{1--12}.
\newblock


\bibitem[Li et~al\mbox{.}(2023b)]%
        {brainstorm}
\bibfield{author}{\bibinfo{person}{Xin{-}Ye Li}, \bibinfo{person}{Jiang{-}Tian Xue}, \bibinfo{person}{Zheng Xie}, {and} \bibinfo{person}{Ming Li}.} \bibinfo{year}{2023}\natexlab{b}.
\newblock \showarticletitle{Think Outside the Code: Brainstorming Boosts Large Language Models in Code Generation}.
\newblock \bibinfo{journal}{\emph{CoRR}}  \bibinfo{volume}{2305.10679} (\bibinfo{year}{2023}), \bibinfo{pages}{1--13}.
\newblock


\bibitem[Li et~al\mbox{.}(2022)]%
        {alphacode}
\bibfield{author}{\bibinfo{person}{Yujia Li}, \bibinfo{person}{David Choi}, \bibinfo{person}{Junyoung Chung}, \bibinfo{person}{Nate Kushman}, \bibinfo{person}{Julian Schrittwieser}, \bibinfo{person}{Rémi Leblond}, \bibinfo{person}{Tom Eccles}, \bibinfo{person}{James Keeling}, \bibinfo{person}{Felix Gimeno}, \bibinfo{person}{Agustin~Dal Lago}, \bibinfo{person}{Thomas Hubert}, \bibinfo{person}{Peter Choy}, \bibinfo{person}{Cyprien de Masson~d’Autume}, \bibinfo{person}{Igor Babuschkin}, \bibinfo{person}{Xinyun Chen}, \bibinfo{person}{Po-Sen Huang}, \bibinfo{person}{Johannes Welbl}, \bibinfo{person}{Sven Gowal}, \bibinfo{person}{Alexey Cherepanov}, \bibinfo{person}{James Molloy}, \bibinfo{person}{Daniel~J. Mankowitz}, \bibinfo{person}{Esme~Sutherland Robson}, \bibinfo{person}{Pushmeet Kohli}, \bibinfo{person}{Nando de Freitas}, \bibinfo{person}{Koray Kavukcuoglu}, {and} \bibinfo{person}{Oriol Vinyals}.} \bibinfo{year}{2022}\natexlab{}.
\newblock \showarticletitle{Competition-Level Code Generation with AlphaCode}.
\newblock \bibinfo{journal}{\emph{Science}}  \bibinfo{volume}{378} (\bibinfo{year}{2022}), \bibinfo{pages}{1092--1097}.
\newblock


\bibitem[Liu et~al\mbox{.}(2023)]%
        {evalplus}
\bibfield{author}{\bibinfo{person}{Jiawei Liu}, \bibinfo{person}{Chunqiu~Steven Xia}, \bibinfo{person}{Yuyao Wang}, {and} \bibinfo{person}{Lingming Zhang}.} \bibinfo{year}{2023}\natexlab{}.
\newblock \showarticletitle{Is Your Code Generated by ChatGPT Really Correct? Rigorous Evaluation of Large Language Models for Code Generation}. In \bibinfo{booktitle}{\emph{NeurIPS}}. \bibinfo{publisher}{Curran Associates, Inc.}, \bibinfo{address}{New Orleans, LA, USA}, \bibinfo{pages}{21558--21572}.
\newblock


\bibitem[Luo et~al\mbox{.}(2024)]%
        {wizardcoder}
\bibfield{author}{\bibinfo{person}{Ziyang Luo}, \bibinfo{person}{Can Xu}, \bibinfo{person}{Pu Zhao}, \bibinfo{person}{Qingfeng Sun}, \bibinfo{person}{Xiubo Geng}, \bibinfo{person}{Wenxiang Hu}, \bibinfo{person}{Chongyang Tao}, \bibinfo{person}{Jing Ma}, \bibinfo{person}{Qingwei Lin}, {and} \bibinfo{person}{Daxin Jiang}.} \bibinfo{year}{2024}\natexlab{}.
\newblock \showarticletitle{WizardCoder: Empowering Code Large Language Models with Evol-Instruct}. In \bibinfo{booktitle}{\emph{ICLR}}. \bibinfo{publisher}{OpenReview.net}, \bibinfo{address}{Kigali, Rwanda}, \bibinfo{pages}{1--21}.
\newblock


\bibitem[Mu et~al\mbox{.}(2023)]%
        {clarifygpt}
\bibfield{author}{\bibinfo{person}{Fangwen Mu}, \bibinfo{person}{Lin Shi}, \bibinfo{person}{Song Wang}, \bibinfo{person}{Zhuohao Yu}, \bibinfo{person}{Binquan Zhang}, \bibinfo{person}{Chenxue Wang}, \bibinfo{person}{Shichao Liu}, {and} \bibinfo{person}{Qing Wang}.} \bibinfo{year}{2023}\natexlab{}.
\newblock \showarticletitle{ClarifyGPT: Empowering LLM-based Code Generation with Intention Clarification}.
\newblock \bibinfo{journal}{\emph{CoRR}}  \bibinfo{volume}{2310.10996} (\bibinfo{year}{2023}), \bibinfo{pages}{1--21}.
\newblock


\bibitem[Ni et~al\mbox{.}(2023)]%
        {lever}
\bibfield{author}{\bibinfo{person}{Ansong Ni}, \bibinfo{person}{Srini Iyer}, \bibinfo{person}{Dragomir Radev}, \bibinfo{person}{Veselin Stoyanov}, \bibinfo{person}{Wen tau Yih}, \bibinfo{person}{Sida Wang}, {and} \bibinfo{person}{Xi~Victoria Lin}.} \bibinfo{year}{2023}\natexlab{}.
\newblock \showarticletitle{LEVER: Learning to Verify Language-to-Code Generation with Execution}. In \bibinfo{booktitle}{\emph{ICML}}. \bibinfo{publisher}{PMLR}, \bibinfo{address}{Honolulu, HI, USA}, \bibinfo{pages}{26106--26128}.
\newblock


\bibitem[Olausson et~al\mbox{.}(2024)]%
        {self_repair}
\bibfield{author}{\bibinfo{person}{Theo~X. Olausson}, \bibinfo{person}{Jeevana~Priya Inala}, \bibinfo{person}{Chenglong Wang}, \bibinfo{person}{Jianfeng Gao}, {and} \bibinfo{person}{Armando Solar-Lezama}.} \bibinfo{year}{2024}\natexlab{}.
\newblock \showarticletitle{Is Self-Repair a Silver Bullet for Code Generation?}. In \bibinfo{booktitle}{\emph{ICLR}}. \bibinfo{publisher}{OpenReview.net}, \bibinfo{address}{Vienna, Austria}, \bibinfo{pages}{1--49}.
\newblock


\bibitem[{OpenAI}(2022)]%
        {chatgpt}
\bibfield{author}{\bibinfo{person}{{OpenAI}}.} \bibinfo{year}{2022}\natexlab{}.
\newblock \bibinfo{title}{Introducing {ChatGPT}}.
\newblock \bibinfo{howpublished}{\url{https://openai.com/index/chatgpt/}}.
\newblock
\newblock
\shownote{Accessed May 30, 2024}.


\bibitem[{OpenAI}(2023)]%
        {gpt4_report}
\bibfield{author}{\bibinfo{person}{{OpenAI}}.} \bibinfo{year}{2023}\natexlab{}.
\newblock \showarticletitle{{GPT-4} Technical Report}.
\newblock \bibinfo{journal}{\emph{CoRR}}  \bibinfo{volume}{2303.08774} (\bibinfo{year}{2023}), \bibinfo{pages}{1--100}.
\newblock


\bibitem[Peng et~al\mbox{.}(2023)]%
        {Peng2023The}
\bibfield{author}{\bibinfo{person}{Sida Peng}, \bibinfo{person}{Eirini Kalliamvakou}, \bibinfo{person}{Peter Cihon}, {and} \bibinfo{person}{Mert Demirer}.} \bibinfo{year}{2023}\natexlab{}.
\newblock \showarticletitle{The Impact of {AI} on Developer Productivity: Evidence from {GitHub} Copilot}.
\newblock \bibinfo{journal}{\emph{CoRR}}  \bibinfo{volume}{2302.06590} (\bibinfo{year}{2023}), \bibinfo{pages}{1--19}.
\newblock


\bibitem[Petersen et~al\mbox{.}(2009)]%
        {waterfull}
\bibfield{author}{\bibinfo{person}{Kai Petersen}, \bibinfo{person}{Claes Wohlin}, {and} \bibinfo{person}{Dejan Baca}.} \bibinfo{year}{2009}\natexlab{}.
\newblock \showarticletitle{The Waterfall Model in Large-Scale Development}. In \bibinfo{booktitle}{\emph{PROFES}}. \bibinfo{publisher}{Springer}, \bibinfo{address}{Oulu, Finland}, \bibinfo{pages}{386--400}.
\newblock


\bibitem[Qian et~al\mbox{.}(2024)]%
        {chatdev}
\bibfield{author}{\bibinfo{person}{Chen Qian}, \bibinfo{person}{Wei Liu}, \bibinfo{person}{Hongzhang Liu}, \bibinfo{person}{Nuo Chen}, \bibinfo{person}{Yufan Dang}, \bibinfo{person}{Jiahao Li}, \bibinfo{person}{Cheng Yang}, \bibinfo{person}{Weize Chen}, \bibinfo{person}{Yusheng Su}, \bibinfo{person}{Xin Cong}, \bibinfo{person}{Juyuan Xu}, \bibinfo{person}{Dahai Li}, \bibinfo{person}{Zhiyuan Liu}, {and} \bibinfo{person}{Maosong Sun}.} \bibinfo{year}{2024}\natexlab{}.
\newblock \showarticletitle{ChatDev: Communicative Agents for Software Development}. In \bibinfo{booktitle}{\emph{ACL}}. \bibinfo{publisher}{ACL}, \bibinfo{address}{Bangkok, Thailand}, \bibinfo{pages}{15174–15186}.
\newblock


\bibitem[Ren et~al\mbox{.}(2023)]%
        {Ren2023}
\bibfield{author}{\bibinfo{person}{Xiaoxue Ren}, \bibinfo{person}{Xinyuan Ye}, \bibinfo{person}{Dehai Zhao}, \bibinfo{person}{Zhenchang Xing}, {and} \bibinfo{person}{Xiaohu Yang}.} \bibinfo{year}{2023}\natexlab{}.
\newblock \showarticletitle{From Misuse to Mastery: Enhancing Code Generation with Knowledge-Driven {AI} Chaining}. In \bibinfo{booktitle}{\emph{ASE}}. \bibinfo{publisher}{IEEE/ACM}, \bibinfo{address}{Luxembourg}, \bibinfo{pages}{976--987}.
\newblock


\bibitem[Ridnik et~al\mbox{.}(2024)]%
        {ridnik2024code}
\bibfield{author}{\bibinfo{person}{Tal Ridnik}, \bibinfo{person}{Dedy Kredo}, {and} \bibinfo{person}{Itamar Friedman}.} \bibinfo{year}{2024}\natexlab{}.
\newblock \showarticletitle{Code Generation with AlphaCodium: From Prompt Engineering to Flow Engineering}.
\newblock \bibinfo{journal}{\emph{CoRR}}  \bibinfo{volume}{2401.08500} (\bibinfo{year}{2024}), \bibinfo{pages}{1--10}.
\newblock


\bibitem[Rozi{\`{e}}re et~al\mbox{.}(2023)]%
        {codellama}
\bibfield{author}{\bibinfo{person}{Baptiste Rozi{\`{e}}re}, \bibinfo{person}{Jonas Gehring}, \bibinfo{person}{Fabian Gloeckle}, \bibinfo{person}{Sten Sootla}, \bibinfo{person}{Itai Gat}, \bibinfo{person}{Xiaoqing~Ellen Tan}, \bibinfo{person}{Yossi Adi}, \bibinfo{person}{Jingyu Liu}, \bibinfo{person}{Tal Remez}, \bibinfo{person}{J{\'{e}}r{\'{e}}my Rapin}, \bibinfo{person}{Artyom Kozhevnikov}, \bibinfo{person}{Ivan Evtimov}, \bibinfo{person}{Joanna Bitton}, \bibinfo{person}{Manish Bhatt}, \bibinfo{person}{Cristian Canton{-}Ferrer}, \bibinfo{person}{Aaron Grattafiori}, \bibinfo{person}{Wenhan Xiong}, \bibinfo{person}{Alexandre D{\'{e}}fossez}, \bibinfo{person}{Jade Copet}, \bibinfo{person}{Faisal Azhar}, \bibinfo{person}{Hugo Touvron}, \bibinfo{person}{Louis Martin}, \bibinfo{person}{Nicolas Usunier}, \bibinfo{person}{Thomas Scialom}, {and} \bibinfo{person}{Gabriel Synnaeve}.} \bibinfo{year}{2023}\natexlab{}.
\newblock \showarticletitle{Code Llama: Open Foundation Models for Code}.
\newblock \bibinfo{journal}{\emph{CoRR}}  \bibinfo{volume}{2308.12950} (\bibinfo{year}{2023}), \bibinfo{pages}{1--48}.
\newblock


\bibitem[Shi et~al\mbox{.}(2022)]%
        {mbr_exe}
\bibfield{author}{\bibinfo{person}{Freda Shi}, \bibinfo{person}{Daniel Fried}, \bibinfo{person}{Marjan Ghazvininejad}, \bibinfo{person}{Luke Zettlemoyer}, {and} \bibinfo{person}{Sida~I. Wang}.} \bibinfo{year}{2022}\natexlab{}.
\newblock \showarticletitle{Natural Language to Code Translation with Execution}. In \bibinfo{booktitle}{\emph{EMNLP}}. \bibinfo{publisher}{ACL}, \bibinfo{address}{Abu Dhabi, United Arab Emirates}, \bibinfo{pages}{3533--3546}.
\newblock


\bibitem[Shinn et~al\mbox{.}(2023)]%
        {reflexion}
\bibfield{author}{\bibinfo{person}{Noah Shinn}, \bibinfo{person}{Federico Cassano}, \bibinfo{person}{Ashwin Gopinath}, \bibinfo{person}{Karthik Narasimhan}, {and} \bibinfo{person}{Shunyu Yao}.} \bibinfo{year}{2023}\natexlab{}.
\newblock \showarticletitle{Reflexion: Language Agents with Verbal Reinforcement Learning}. In \bibinfo{booktitle}{\emph{NeurIPS}}, Vol.~\bibinfo{volume}{36}. \bibinfo{publisher}{Curran Associates, Inc.}, \bibinfo{address}{New Orleans, LA, USA}, \bibinfo{pages}{8634--8652}.
\newblock


\bibitem[Su et~al\mbox{.}(2024)]%
        {Su2024ARKS}
\bibfield{author}{\bibinfo{person}{Hongjin Su}, \bibinfo{person}{Shuyang Jiang}, \bibinfo{person}{Yuhang Lai}, \bibinfo{person}{Haoyuan Wu}, \bibinfo{person}{Boao Shi}, \bibinfo{person}{Che Liu}, \bibinfo{person}{Qian Liu}, {and} \bibinfo{person}{Tao Yu}.} \bibinfo{year}{2024}\natexlab{}.
\newblock \showarticletitle{{ARKS:} Active Retrieval in Knowledge Soup for Code Generation}.
\newblock \bibinfo{journal}{\emph{CoRR}}  \bibinfo{volume}{2402.12317} (\bibinfo{year}{2024}), \bibinfo{pages}{1--16}.
\newblock


\bibitem[Valmeekam et~al\mbox{.}(2023)]%
        {Valmeekam2023Can}
\bibfield{author}{\bibinfo{person}{Karthik Valmeekam}, \bibinfo{person}{Matthew Marquez}, {and} \bibinfo{person}{Subbarao Kambhampati}.} \bibinfo{year}{2023}\natexlab{}.
\newblock \showarticletitle{Can Large Language Models Really Improve by Self-critiquing Their Own Plans?}
\newblock \bibinfo{journal}{\emph{CoRR}}  \bibinfo{volume}{2310.08118} (\bibinfo{year}{2023}), \bibinfo{pages}{1--6}.
\newblock


\bibitem[Wang et~al\mbox{.}(2024)]%
        {intervenor}
\bibfield{author}{\bibinfo{person}{Hanbin Wang}, \bibinfo{person}{Zhenghao Liu}, \bibinfo{person}{Shuo Wang}, \bibinfo{person}{Ganqu Cui}, \bibinfo{person}{Ning Ding}, \bibinfo{person}{Zhiyuan Liu}, {and} \bibinfo{person}{Ge Yu}.} \bibinfo{year}{2024}\natexlab{}.
\newblock \showarticletitle{{INTERVENOR:} Prompt the Coding Ability of Large Language Models with the Interactive Chain of Repairing}. In \bibinfo{booktitle}{\emph{Findings of ACL}}. \bibinfo{publisher}{ACL}, \bibinfo{address}{Bangkok, Thailand}, \bibinfo{pages}{2081–2107}.
\newblock


\bibitem[Wei et~al\mbox{.}(2022)]%
        {cot}
\bibfield{author}{\bibinfo{person}{Jason Wei}, \bibinfo{person}{Xuezhi Wang}, \bibinfo{person}{Dale Schuurmans}, \bibinfo{person}{Maarten Bosma}, \bibinfo{person}{brian ichter}, \bibinfo{person}{Fei Xia}, \bibinfo{person}{Ed Chi}, \bibinfo{person}{Quoc~V Le}, {and} \bibinfo{person}{Denny Zhou}.} \bibinfo{year}{2022}\natexlab{}.
\newblock \showarticletitle{Chain-of-Thought Prompting Elicits Reasoning in Large Language Models}. In \bibinfo{booktitle}{\emph{NeurIPS}}. \bibinfo{publisher}{Curran Associates, Inc.}, \bibinfo{address}{New Orleans, LA, USA}, \bibinfo{pages}{24824--24837}.
\newblock


\bibitem[Williams(2001)]%
        {pair_programming}
\bibfield{author}{\bibinfo{person}{L. Williams}.} \bibinfo{year}{2001}\natexlab{}.
\newblock \showarticletitle{Integrating pair programming into a software development process}. In \bibinfo{booktitle}{\emph{14th Conference on Software Engineering Education and Training. `In search of a software engineering profession' (Cat. No. PR01059)}}. \bibinfo{publisher}{IEEE}, \bibinfo{address}{Charlotte, NC, USA}, \bibinfo{pages}{27--36}.
\newblock


\bibitem[Yang et~al\mbox{.}(2024)]%
        {Yang2024Exploring}
\bibfield{author}{\bibinfo{person}{Zhen Yang}, \bibinfo{person}{Fang Liu}, \bibinfo{person}{Zhongxing Yu}, \bibinfo{person}{Jacky~Wai Keung}, \bibinfo{person}{Jia Li}, \bibinfo{person}{Shuo Liu}, \bibinfo{person}{Yifan Hong}, \bibinfo{person}{Xiaoxue Ma}, \bibinfo{person}{Zhi Jin}, {and} \bibinfo{person}{Ge Li}.} \bibinfo{year}{2024}\natexlab{}.
\newblock \showarticletitle{Exploring and Unleashing the Power of Large Language Models in Automated Code Translation}.
\newblock \bibinfo{journal}{\emph{Proc. {ACM} Softw. Eng.}} \bibinfo{volume}{1}, \bibinfo{number}{{FSE}} (\bibinfo{year}{2024}), \bibinfo{pages}{1585--1608}.
\newblock


\bibitem[Yu et~al\mbox{.}(2024)]%
        {Yu2024CoderEval}
\bibfield{author}{\bibinfo{person}{Hao Yu}, \bibinfo{person}{Bo Shen}, \bibinfo{person}{Dezhi Ran}, \bibinfo{person}{Jiaxin Zhang}, \bibinfo{person}{Qi Zhang}, \bibinfo{person}{Yuchi Ma}, \bibinfo{person}{Guangtai Liang}, \bibinfo{person}{Ying Li}, \bibinfo{person}{Qianxiang Wang}, {and} \bibinfo{person}{Tao Xie}.} \bibinfo{year}{2024}\natexlab{}.
\newblock \showarticletitle{{CoderEval}: A Benchmark of Pragmatic Code Generation with Generative Pre-trained Models}. In \bibinfo{booktitle}{\emph{ICSE}}. \bibinfo{publisher}{ACM}, \bibinfo{address}{Lisbon, Portugal}, \bibinfo{pages}{37:1--37:12}.
\newblock


\bibitem[Zhang et~al\mbox{.}(2023a)]%
        {self_edit}
\bibfield{author}{\bibinfo{person}{Kechi Zhang}, \bibinfo{person}{Zhuo Li}, \bibinfo{person}{Jia Li}, \bibinfo{person}{Ge Li}, {and} \bibinfo{person}{Zhi Jin}.} \bibinfo{year}{2023}\natexlab{a}.
\newblock \showarticletitle{Self-Edit: Fault-Aware Code Editor for Code Generation}. In \bibinfo{booktitle}{\emph{ACL}}. \bibinfo{publisher}{ACL}, \bibinfo{address}{Toronto, Canada}, \bibinfo{pages}{769--787}.
\newblock


\bibitem[Zhang et~al\mbox{.}(2023b)]%
        {zhang2023algo}
\bibfield{author}{\bibinfo{person}{Kexun Zhang}, \bibinfo{person}{Danqing Wang}, \bibinfo{person}{Jingtao Xia}, \bibinfo{person}{William~Yang Wang}, {and} \bibinfo{person}{Lei Li}.} \bibinfo{year}{2023}\natexlab{b}.
\newblock \showarticletitle{ALGO: Synthesizing Algorithmic Programs with LLM-generated Oracle Verifiers}. In \bibinfo{booktitle}{\emph{NeurIPS}}. \bibinfo{publisher}{Curran Associates Inc.}, \bibinfo{address}{New Orleans, LA, USA}, \bibinfo{pages}{54769–54784}.
\newblock


\bibitem[Zhang et~al\mbox{.}(2023c)]%
        {codereviewer}
\bibfield{author}{\bibinfo{person}{Tianyi Zhang}, \bibinfo{person}{Tao Yu}, \bibinfo{person}{Tatsunori Hashimoto}, \bibinfo{person}{Mike Lewis}, \bibinfo{person}{Wen{-}Tau Yih}, \bibinfo{person}{Daniel Fried}, {and} \bibinfo{person}{Sida Wang}.} \bibinfo{year}{2023}\natexlab{c}.
\newblock \showarticletitle{Coder Reviewer Reranking for Code Generation}. In \bibinfo{booktitle}{\emph{ICML}}, Vol.~\bibinfo{volume}{202}. \bibinfo{publisher}{PMLR}, \bibinfo{address}{Honolulu, HI, USA}, \bibinfo{pages}{41832--41846}.
\newblock


\bibitem[Zheng et~al\mbox{.}(2023)]%
        {Zheng2023CodeGeeX}
\bibfield{author}{\bibinfo{person}{Qinkai Zheng}, \bibinfo{person}{Xiao Xia}, \bibinfo{person}{Xu Zou}, \bibinfo{person}{Yuxiao Dong}, \bibinfo{person}{Shan Wang}, \bibinfo{person}{Yufei Xue}, \bibinfo{person}{Lei Shen}, \bibinfo{person}{Zihan Wang}, \bibinfo{person}{Andi Wang}, \bibinfo{person}{Yang Li}, \bibinfo{person}{Teng Su}, \bibinfo{person}{Zhilin Yang}, {and} \bibinfo{person}{Jie Tang}.} \bibinfo{year}{2023}\natexlab{}.
\newblock \showarticletitle{{CodeGeeX}: {A} Pre-Trained Model for Code Generation with Multilingual Benchmarking on {HumanEval-X}}. In \bibinfo{booktitle}{\emph{KDD}}. \bibinfo{publisher}{ACM}, \bibinfo{address}{Long Beach, CA, USA}, \bibinfo{pages}{5673--5684}.
\newblock


\bibitem[Zhong et~al\mbox{.}(2024)]%
        {ldb}
\bibfield{author}{\bibinfo{person}{Li Zhong}, \bibinfo{person}{Zilong Wang}, {and} \bibinfo{person}{Jingbo Shang}.} \bibinfo{year}{2024}\natexlab{}.
\newblock \showarticletitle{Debug like a Human: A Large Language Model Debugger via Verifying Runtime Execution Step by Step}. In \bibinfo{booktitle}{\emph{Findings of ACL}}. \bibinfo{publisher}{ACL}, \bibinfo{address}{Bangkok, Thailand}, \bibinfo{pages}{851–870}.
\newblock


\end{thebibliography}


\end{document}